\documentclass[aps,prb,groupedaddress,showkeys,showpacs,amsmath]{revtex4}

\usepackage{amssymb}
\usepackage{amsmath}
\usepackage{amscd}
\usepackage{graphicx}
\usepackage{amsbsy}
\usepackage{color}
\graphicspath{{./FIGS/}}

\usepackage{tocloft}

\usepackage{times}

\usepackage{titlesec}
\titleformat{\section}{\normalsize\bfseries}{\thesection}{0.5em}{}
\titleformat{\subsection}{\normalsize\bfseries}{\thesubsection}{0.5em}{}

\begin{document}

\title{Theoretical calculations of phase diagrams and self-assembly in patchy colloids}

\author{Achille Giacometti, Flavio Romano and Francesco Sciortino}
\email{achille.giacometti@unive.it}
\affiliation{Dipartimento di Dipartimento di Scienze Molecolari e Nanosistemi, Universit\`a Ca' Foscari Venezia,
Calle Larga S. Marta DD2137, I-30123 Venezia, Italy}

\author{Flavio Romano}
\email{flavio.romano@gmail.com}
\affiliation{Physical and Theoretical Chemistry Laboratory, Oxford University (UK)}
\date{\today}

\author{Francesco Sciortino}
\email{francesco.sciortino@uniroma1.it}
\affiliation{Dipartimento di Fisica and
  CNR-ISC,  {\it Sapienza} Universit\`a di Roma, Piazzale A. Moro 5, 00185 Roma, Italy}


\maketitle

\newpage



\newpage


\section{Introduction}
\label{sec:intro}
Self-assembly processes in patchy colloids represent one of the most
striking examples where experimental methodologies and theoretical tools
have progressed in parallel within a relatively short time scale~\cite{Whitesides02,Glotzer04,Glotzer07}. While the former have been
addressed elsewhere in this volume and in recent reviews
\cite{Walther09,Pawar10}, in this contribution we will address the latter
and, more specifically, the main methodologies that have been envisaged
over the years to tackle the computation of the phase diagrams and phase
transitions from one phase to another in dispersions of new-generation
colloids, i.e., particles interacting via non-spherical potentials. Indeed,
chemists and material scientists are starting to gain control on the
shapes\cite{Blaad_03} and local properties of colloids. Hard cubes,
tetrahedra, cones, rods as well as composed shapes of nano or microscopic
size have made their appearance in the laboratories, and are becoming 
available in bulk quantities. Patterning of the surface properties of these
particles~\cite{Manoh_03,Cho_05,kegel} provides additional degrees of freedom
to be exploited by scientists to engineer materials with peculiar
properties. Patches on the particle surface can be functionalized with
specific molecules\cite{hiddessen1,mohovald} (including DNA single
strands\cite{dna,Milam03La}) to create hydrophobic or hydrophilic areas,
providing specificity to the particle-particle
interaction~\cite{Pawar10,Bianchi11}.

Statistical physics provides a very rich and flexible toolbox to study
thermo-physical and structural properties of complex
fluids\cite{Hansen86,Gray84}, especially when coupled with the most recent
and powerful computing techniques devised to deal with systems with
a large number of degrees of freedom~\cite{Frenkel02,Allen87}. While simple
liquids and conventional colloidal systems have a long and venerable
tradition~\cite{Lyklema91}, theoretical studies on patchy colloids are
relatively new, as in the past it was always tacitly assumed that even
the unavoidable inhomogeneities in their surface composition could be
neglected at a sufficiently coarse-grained scale. This is not the case,
however, for patchy colloids that have surface patterns, chemical
compositions and functionalities that are explicitly meant to be
inhomogeneous~\cite{Walther09,Pawar10,Bianchi11}. Hence, the corresponding
pair potentials describing inter-particle interactions depend on their
relative orientations besides distances, and the analysis clearly becomes
more complex. This is, however, by no means an insurmountable difficulty,
as several analytical and computational techniques have been devised
in statistical physics to cope with the orientational dependence of the
potentials~\cite{Hansen86,Gray84}.

In this Chapter, we shall discuss some of them in the framework of a particular pair potential that can be reckoned as a reasonable compromise between the complexity of the real interactions, and the necessary simplicity required to keep the analysis amenable. The basic idea of the model is built on the hard-sphere model, by providing 
 a fraction of the surface sphere with a square-well character. This attractive region can be either condensed into a single large patch, or distributed over two (or more) patches symmetrically placed over the surface. Different spheres then interact
via a square-well or hard-sphere potentials depending on their relative orientations and distances.

This model was proposed in 2003 by Kern and Frenkel \cite{Kern03} and has ever since attracted considerable attention.  There are two main reasons for this. On the one hand, the model is very flexible, as both the size and the number of the patches can be independently
tuned, and this allows to mimic several different physical situations ranging from nanocolloids with more isolated attractive spots\cite{Bianchi06} 
to globular proteins with large regions of solvophobic exposed surfaces \cite{Liu07,Gogelein08}.  On the other hand, the phase diagram obtained from the model can be directly compared with those obtained from experiments, as recently shown in several cases \cite{Hong08,Chen11,Romano11_a,Romano11_b}. 
In addition, the model displays some unusual features that can be paradigmatic for more complex systems \cite{Sciortino09,prl-lisbona,lisbona_lungo,Sciortino10}.

The aim of this Chapter is to introduce the main theoretical techniques to the evaluation of the phase diagram. This includes various 
Monte Carlo techniques (Section \ref{sec:montecarlo}), integral equations (Section \ref{sec:integral}), and perturbation theory
(Section \ref{sec:perturbation}). The level is intended to be pedagogic, with the main ideas behind each method
outlined for non-experts in the field. Emphasis will be given on the calculations of  thermodynamic quantities necessary for the phase diagram analysis, and hence a number of additional important results related to structural properties and other thermodynamical probes  will be omitted. 
\section{The Kern-Frenkel model}
\label{sec:model}
Consider a set of $N$ identical hard-spheres of diameter $\sigma$ in a volume $V$ at temperature $T$ suspended in a microscopic fluid.
When the surface of the spheres are uniform with no other interactions as their steric hindrance, the model has been often
employed as a paradigm of sterically stabilized colloidal suspensions in the limit of high temperature or good solvent.

As discussed elsewhere in this volume, colloids that are envisaged as elementary building blocks for the self-assembly
process, are patchy colloids with different philicities in different part of the surface \cite{Walther09,Pawar10}. 
This means, for instance, that one fraction of the
surface may be solvophilic and the other solvophobic. In solution, the solvophobic part will tend to avoid contact with the solvent
and hence will act as an effective attractive force in the presence of another solvophobic patch laying on a different sphere.

One can then consider the following model that  was introduced in 2003 by Kern and Frenkel in the present form \cite{Kern03}, but it is worth remarking 
that the idea  of considering hard spheres
and decorating them with patches of various forms and patterns dates back to much earlier, and several earlier versions in different fields
can be considered as its ancestors \cite{bal,depablo,nezbeda,Lomakin99}.

\begin{figure}[t]
\begin{center}
\includegraphics[width=8.5cm]{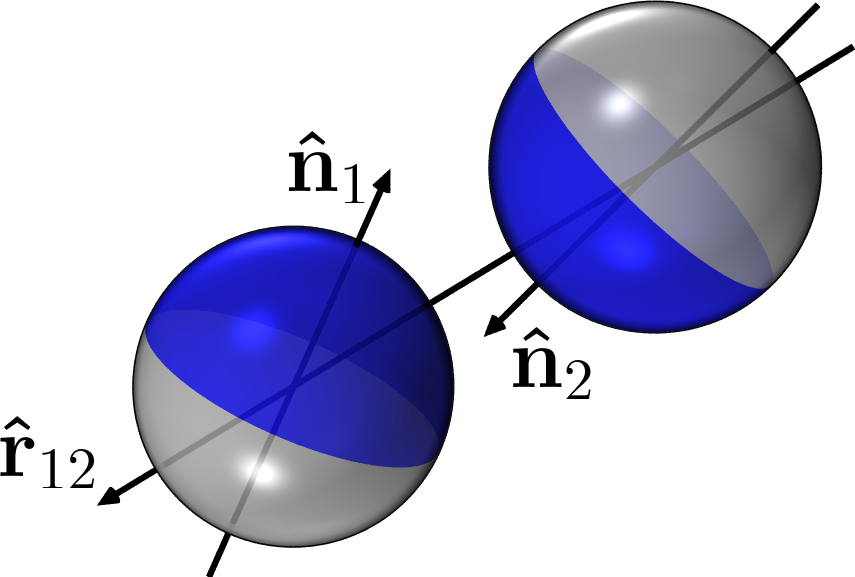}
\end{center}
\caption{Sketch of one-patch parchy particles as modelled by the
Kern-Frenkel potential. The surface of each sphere is partitioned into an
attractive part (color code: blue) and a repulsive part (color code:
white). The unit vectors $\hat{\mathbf{n}}_{1}$ and $\hat{\mathbf{n}}_{2}$
define the orientations of the pathces, whereas the vector
$\hat{\mathbf{r}}_{12}$ joins the centers of the two spheres, from sphere
$1$ to sphere $2$. The particular case shown corresponds to a $50\%$
fraction of attractive surface (coverage $\chi=0.5$).
\label{fig:fig1}}
\end{figure}

A circular patch is attached to the surface of each sphere, as depicted in Figure \ref{fig:fig1}, with the central position of the patch identified by the unit vector  $\hat{\textbf{n}}$, and  its amplitude measured by the angle $\theta_0$. Unlike the case of uniform sphere the interactions among spheres is anisotropic as it depends on the relative orientation of the unit vectors on each spheres with the direction connecting their centers.
Then the idea is that two spheres attract each other if they are within the range of the attractive potential, with the corresponding attractive
patches on each sphere properly aligned.

If $\hat{\textbf{n}}_{1}$ and  $\hat{\textbf{n}}_{2}$ are the unit vectors associated with each patch on spheres $1$ and $2$, 
and  $\hat{\mathbf{r}}_{12}$ is the direction connecting the centers of the two spheres, then the interparticle potential reads
\begin{eqnarray}
\Phi\left(12\right) &=& \phi_{0} \left(r_{12}\right) +\Phi_{\text{I}}\left(12\right)\,\mbox{,}
\label{kf:eq1}
\end{eqnarray}
where the first term is the hard-sphere contribution
\begin{equation}
\phi_{0}\left(r\right)= \left\{ 
\begin{array}{ccc}
\infty,    &  &   0<r< \sigma    \\ 
0,          &  &   \sigma < r \ %
\end{array}%
\right.\,\mbox{,} \label{kf:eq1b}
\end{equation}
and the second term
\begin{eqnarray}
\Phi_{\text{I}} \left(\hat{\mathbf{n}}_1,
\hat{\mathbf{n}}_2,{\mathbf{r}}_{12} \right)&=& \phi_{\text{SW}} \left(r_{12}\right) \Psi\left(\hat{\mathbf{n}}_1,
\hat{\mathbf{n}}_2,\hat{\mathbf{r}}_{12} \right)
\label{kf:eq1c}
\end{eqnarray}
is the orientation-dependent attractive part which can be factorized into an isotropic square-well tail
\begin{equation}
\phi_{\text{SW}}\left(r\right)= \left\{ 
\begin{array}{ccc}
- \epsilon, &  &   \sigma<r< \lambda \sigma   \\ 
0,          &  &   \lambda \sigma < r \ %
\end{array}%
\right.\,\mbox{,} \label{kf:eq2}
\end{equation}
multiplied by an angular dependent factor
\begin{equation}
\Psi\left(\hat{\mathbf{n}}_1,
\hat{\mathbf{n}}_2,\hat{\mathbf{r}}_{12}\right)= \left\{ 
\begin{array}{ccccc}
1,    & \text{if}  &   \hat{\mathbf{n}}_1 \cdot \hat{\mathbf{r}}_{12} \ge \cos \theta_0 & \text{and} &  
-\hat{\mathbf{n}}_2 \cdot \hat{\mathbf{r}}_{12} \ge \cos \theta_0 \\ 
0,    &  & &\text{otherwise} & \
\end{array}%
\right.\,\mbox{.} \label{kf:eq3}
\end{equation}
Here $\sigma$ is the sphere diameter, $(\lambda-1) \sigma$ is the width of
the square-well interaction and $\epsilon$ its depth. $2 \theta_0$ defines
the angular amplitude of the patch. The unit vectors
$\hat{\mathbf{n}}_{i}(\omega_{i})$, ($i=1,2$), are defined by the spherical
angles $\omega_i=(\theta_i,\varphi_i)$ in an arbitrarily oriented
coordinate frame and $\hat{\mathbf{r}}_{12}(\Omega)$ is identified by the
spherical angle $\Omega$ in the same frame. Reduced units, for temperature
$T^*=k_B T/\epsilon$, pressure $P^{*}=\beta P/\rho$ and density
$\rho^*=\rho \sigma^3$, will be used throughout, with $k_B$ being the
Boltzmann constant. For future reference, we also introduce the packing
fraction $\eta=\pi \rho^{*}/6$. 

The coverage $\chi$ is the fraction of attractive surface on the particle.
$\chi$ can be related to the patch half-width $\theta_0$  as
\begin{eqnarray}
\label{kf:eq4}
\chi &=& \left\langle \Psi\left(\hat{\mathbf{n}}_1,\hat{\mathbf{n}}_2,\hat{\mathbf{\Omega}} 
\right) \right \rangle_{\omega_1 \omega_2}^{1/2} = \frac{1-\cos \theta_0 }{2}
\end{eqnarray}
where we have introduced the angular average
\begin{eqnarray}
\label{kf:eq5}
\left \langle \ldots \right \rangle_{\omega} &=& \frac{1}{4 \pi} \int d \omega \ldots\,\mbox{.}
\end{eqnarray}
\section{The tools of statistical physics}
\label{sec:tools}
Statistical physics has developed a number of different theoretical approaches to compute the thermophysical
properties of a fluid \cite{Hansen86,Gray84,Frenkel02}. In order to compare with experiments, we are most interested in the computation of the
fluid-fluid and fluid-solid phase diagram on the one hand, and on the specific mechanism driving aggregation,
and hence self-assembly, on the other hand. Among this arsenal of different available techniques, here we will review three different methodologies that were
recently exploited in the framework of the Kern-Frenkel model. These are Monte Carlo simulations \cite{Frenkel02,Allen87}, integral equation theories
\cite{Hansen86} and thermodynamical perturbation theories~\cite{Gray84}.

Monte Carlo simulations are undoubtably one the most efficient ways to
accurately compute the properties of a model fluid. As discussed in more
details below, the main limitations of simulations are that they can be
very demanding from a computational point of view, especially for
sufficiently realistic potentials, and that they are unable to distinguish
metastable from stable equilibrium states. On the other hand, they provide
virtually exact estimates of all static quantities of interest. Several
improvements have been proposed over the years, some of them triggered by
the problems discussed in this contribution, so that the methodology is 
very well established and by now extensively reviewed and described in
detail in several books (see Refs.~\onlinecite{Frenkel02,Allen87} and
references therein). The case of patchy colloids, however, is relatively
recent, although it builds upon previous established procedures on other
complex fluids.

Integral equation and thermodynamics perturbation theories are two of the
main methodologies from the toolbox of Statistical Physics that are at the
basis of our current understaing of simple and molecular
fluids~\cite{Hansen86,Gray84}. In spite of their known drawbacks and
shortcomings, they are known to provide reliable predictions for both
structural and thermodynamical properties each in their own domain of
applications. Their applications to patchy colloids is a rather natural,
albeit not trivial, extension of formalisms already developed in the last
two decades for molecular fluids \cite{Gray84}. As it will become clear,
they both become particularly attractive in view of the large computational
effort involved in Monte Carlo simulations. In addition, they are able to
access to some details and nuances that are not easily accessible by other
methods.

\section{Monte Carlo simulations}
\label{sec:montecarlo}
The aim of Monte Carlo simulations is the computation of thermodynamic
quantities by performing an average over a suitable ensemble of
microstates. The choice of the ensemble is dictated both by the quantities
to be computed and by the specific system under investigation, for which
one ensemble can be more convenient than the others. Below we review the
most interesting techniques that have been used to calculate phase diagram
of patchy colloid models.

\subsection{Canonical $NVT$ and $NPT$ methods}
\label{subsec:nvt}
Simulations in the $NVT$ (isothermal-isochoric) and $NPT$
(isothermal-isobaric) ensembles are probably the most common example.
In these ensembles, the
number of particles $N$, the temperature $T$ and the volume $V$ ($NVT$) or
the pressure $P$ ($NPT$) are held constant. The Markov chain in
configuration space is constructed via a sequence of translational and
rotational moves, accepted with an appropriate probability that depends upon
the change in potential energy and $T$. In the $NPT$ case, the volume is
also varied. With a proper choice of the acceptance probability, the system
first evolves toward equilibrium and then starts to sample equilibrium
configurations with the Boltzmann statistical weight. The equilibration
process can be rather long, especially in cases where kinetic traps are
present (as in the vicinity of gel or glass transitions) or when an
activation barrier needs to be overcome. This last case arises when the
system is metastable with respect to a lower energy phase or when it
organizes into mesoscopic structures and specific self-assembly
processes involving large numbers of particles are requested. The approach
to equilibrium can be monitored by focusing on the time evolution of
collective properties (e.g. the potential energy, the density, the
pressure). Since equilibration can be rather slow, it is highly recommended
to make use of a logarithmic time scale when searching for a drift in the
time dependence of the investigated property.

When a sufficiently large number of statistically independent equilibrium
configurations have been generated and stored, all possible structural
(static) information can be calculated. Typical quantities that are
computed are the total energy $U$, the radial distribution function $g(r)$
and the structure factor. In the case of anisotropic systems, such as
patchy particles, the orientational dependence of the structural properties
needs to be evaluated as well. The center-center $g(r)$ is indeed not
sufficient to evaluate the average potential energy or the pressure, at
variance of the isotropic case where calculation of $U$ or $P$ from the
$g(r)$ usually simply requires a one-dimensional integration.
 
In the case of hard bodies or in the presence of step-wise potentials (e.g.
the square-well potential), direct evaluation of $P$ in $NVT$ simulations
is in principle possible~\cite{eppenga,articoloinpablo} but not
straightforward. To evaluate the equation of state, i.e., the relation
between density and $P$ at constant $T$, the $NPT$ ensemble is often
preferred in this case.

Various additional improvements can be (and are) used to improve the
convergence of the scheme in a way that will be described in each specific
example.

\subsection{Gibbs ensemble method}
\label{subsec:gibbs}
A convenient scheme was devised by
Panagiotopulos~\cite{Panagiotopulos87} to specifically address the problem
of a direct evaluation of the gas-liquid phase coexistence by Monte Carlo
simulations. This is known as the Gibbs Ensemble Monte Carlo (GEMC) method.
$N$ particles are partitioned into two distinct simulation
boxes. In addition to intra-box translational and rotational moves,
particle and volume swap moves (keeping both the total number of particles
and the total volume fixed) are proposed and accepted with the appropriate 
probability~\cite{Frenkel02}. In this way the two coexisting
phases are simulated without the intervention of a interface between them.
When convergence is reached, the densities in the two boxes provide the
value of the coexisting densities of the liquid and gas phase. It should be
pointed out that the GEMC method becomes inefficient when the density of
the liquid phase becomes large, since the probability of inserting a
particle with in a favourable state --- i.e., not overlapping with any
other --- becomes extremely small.

For the specific case of KF particles discussed later in this Chapter, GEMC
simulations have been performed for a system of $1200$ particles in a total
volume of $(16\sigma)^3$. On average, the code attempts one volume change
every five particle-swap moves and $500$ displacement moves. Each
displacement move is composed of a simultaneous random translation of the
particle center (uniformly distributed between $\pm 0.05 \sigma$) and a
rotation (with an angle uniformly distributed between $\pm 0.1$ radians)
around a random axis.

\subsection{Grand-canonical ensemble $\mu VT$}
\label{subsec:grand}
In the neighborhood of the gas-liquid critical point the free-energy
barrier separating the two phases becomes comparable to the amplitude of
the thermal fluctuations of of the relatively small systems that can be
accessed in simulation. In this case, the GEMC method cannot be used to
investigate the system since it becomes size effects and spontaneous
swapping of the average densities between the two boxes.

A precise evaluation of the critical parameters (density and temperature)
can be obtained performing simulations in the grand-canonical $\mu VT$
ensemble~\cite{Frenkel02}, where density fluctuations are accounted for at
fixed volumes and temperature, coupled with the finite-size scaling
analysis envisaged by Wilding~\cite{Wilding95}. MC simulations in the gran
canonical ensemble are implemented by performing 
trial insertions and deletions of particles, besides trial displacements
and rotations. The critical parameters of the system can be extracted by matching the calculated distribution of density fluctuations to the expected
distribution at the critical point, a feature which is largely system
independent~\cite{Wilding95}.

In the implementation of the grand-canonical simulations to the KF model
reported later on, one insertion/deletion attempt was performed, on
average, every $500$ trial translational/rotational displacements. 

\subsection{Fluid-Solid coexistence: thermodynamic integration}
\label{subsec:thermodynamic}
To compute numerically the free energies of the fluid and the crystals and
their coexistence lines it is possible to resort to thermodynamic
integration methods. Details of this procedure were recently given in a
detailed review by Vega et al \cite{Vega08}.

The starting point of the procedure requires the identification of a state
point in the pressure-temperature plane where two phases, I and II, share
the same chemical potential, i.e., $\mu_{\rm I} (P,T) = \mu_{\rm
II}(P,T)\:$. The chemical potential of the fluid can be computed by
thermodynamic integration using the ideal gas as a reference state, and by
integrating the equation of state, $P(\rho)$, at fixed temperature 
\begin{eqnarray}
\label{thermodynamic:eq1}
\frac{\beta F\left(T,\rho\right)}{N}  \, &=& \, \log\left(\rho \sigma^3\right)-1 + \int_{0}^{\rho}{\dfrac{\beta
P/\rho^\prime-1}{\rho^\prime}{\rm d}\rho^\prime}\,
\end{eqnarray}
where $F/N$ is the Helmholtz energy per particle. The first term on the
right-hand-side of Eq.(\ref{thermodynamic:eq1}) is the ideal gas part and
depends upon the system dimensionality. The chemical potential can then be
recovered as
\begin{eqnarray}
\label{thermodynamic:eq2}
\beta\mu \left(P\left(\rho\right),T\right)\, &=& \, \ \frac{\beta F\left( P\left(\rho\right),T\right) }{N}+ \beta P\left(\rho\right) / \rho \:.
\end{eqnarray}
To compute the chemical potential of a crystal one can perform
thermodynamic integration at fixed density and $T$ using an ideal Einstein
crystal as the reference system. This method, known as Frenkel-Ladd
procedure \cite{Frenkel02,Vega08}, is very efficient and by now standard.
Integration of the crystal equation of state provides a way to evaluate the
chemical potential at different $T$ and $P$. The pressure at which the
chemical potential of the fluid and of the crystal are identical along an
isotherm provides the coexisting pressure at the selected $T$.

Starting from a coexistence point, coexistence lines can finally be
inferred by using Gibbs-Duhem integration, as described by Kofke
\cite{Kofke93}, numerically integrating over the Clausius-Clapeyron
equation.

\section{Integral equation theories}
\label{sec:integral}
\subsection{General scheme}
\label{subsec:general}
At first, let us consider simple fluids first where the particles can be
regarded as spherically symmetric. All thermodynamic properties of such
fluids can be straightforwardly computed from the radial distribution
function $g(r)$. In integral equation theory~\cite{Hansen86}, the strategy
to infer the thermophysical properties of a fluid hinges on the calculation
of the total correlation function $h(r) \equiv g(r) - 1$ that, in turn, is
related to the direct correlation function $c(r)$ by the Ornstein-Zernike
(OZ) equation
\begin{eqnarray}
\label{integral:eq1}
h\left( r\right) =c\left( r\right) +\rho \int d \mathbf{r}^{\prime }~c\left( r^{\prime }\right)
~h\left( |\mathbf{r-r}^{\prime }|\right) 
\end{eqnarray}%
Once that $h(r)$ is known, all statistical and thermodynamical properties
can in principle be computed. However, as $h(r)$ depends upon the unknown
quantity $c(r)$, an additional equation involving both quantities is
required for the solution. Unlike equation (\ref{integral:eq1}), which is
exact, the second equation always involves some approximation. This gives
rise to some well known thermodynamic inconsistencies, that are the main
shortcomings of this method, and that may severely limit its applicability.

The second relation between $h(r)$ and $c(r)$ also involves the two-body
potential $\phi(r)$, and can be cast in the general form
\begin{eqnarray}
\label{integral:eq2}
c\left(r\right) &=& \exp{\left[-\beta \phi\left(r\right) + \gamma\left(r\right)+B \left(r\right) \right]}-1 -\gamma\left(r\right)
 \end{eqnarray}
where $\gamma\left( r\right) =h\left( r\right)-c\left( r\right)$ is an
auxiliary function. Although this equation is again exact in principle, it
involves the bridge function $B(r)$ that in general depends upon higher
body correlation functions, so in practice an approximation (closure) is
always necessary. The quality of the results obtained will depend crucially
on the reliability of the involved approximations; several closures have
been proposed over the years with their pros and cons well classified and
under control. Among them, the reference hyper-netted chain (RHNC) stands
out as an optimal trade-off between simplicity and precision of its
predictions, and this is the one that will be the object of the present
Chapter.

Having closed the systems of two equations in two unknowns ($h(r)$, and
$c(r)$), the system may then be solved iteratively with the convolution
appearing in the OZ equation (\ref{integral:eq1}) simplified in Fourier
space, as 
\begin{eqnarray}
\label{integral:eq3}
\hat{h}\left(k\right) &=& \frac{\hat{c}\left(k\right)}{1-\rho \hat{c}\left(k\right)}
\end{eqnarray} 
$\hat{h}(k)$ and $\hat{c}(k)$ being the Fourier transform of $h(r)$ and
$c(r)$ respectively. 

The RHNC closure was introduced by Lado~\cite{Lado73} for spherical
potentials and later extended to molecular fluids~\cite{Lado82a,Lado82b}.
In the RHNC closure, one replaces the exact $B(r)$ appearing in
(\ref{integral:eq2}) by its hard-sphere counterpart $B_0(r)$, that is the
only system for which a reliable expression (the Verlet-Weiss
expression~\cite{Verlet72}) is available. Rosenfeld and
Ashcroft~\cite{Rosenfeld79} demonstrated that the effectiveness of the
reference system could be magnified by treating its parameters as variables
to be optimized in some fashion. It is in fact possible to determine them
via a variational free energy principle~\cite{Lado82} that enhances
internal thermodynamic consistency. With the effective hard sphere diameter
$\sigma_0$ suitably chosen in this way, the RHNC has been shown to provide
rather precise estimates of the chemical potential and pressure, that is
the two crucial thermodynamical quantities needed for the calculation of
phase diagrams.

The case of anisotropic potentials is significantly more complex from the
algorithmic point of view, but the philosophy behind the methodology is
identical. The procedure hinges on a remarkable piece of work carried out
by Fred Lado in a series of papers~\cite{Lado82,Lado82a,Lado82b} in the
framework of molecular fluids and more recently adapted to the case of
patchy colloids. Here we just sketch the idea, referring to
Refs.~\onlinecite{Giacometti09a,Giacometti09b,Giacometti10} for details.

The angular dependent counterparts of Eqs.(\ref{integral:eq1}) and
(\ref{integral:eq2}) are given in terms of $\gamma(12)=h(12)-c(12)$, and
are
\begin{eqnarray}
\label{integral:eq4}
\gamma\left(12\right) &=& \frac{\rho}{4\pi} \int d\mathbf{r}_3 d\omega_3 \left[ \gamma\left(13\right)+c\left(13\right)\right] 
c\left(32\right),
\end{eqnarray}
for the OZ equation, and
\begin{eqnarray}
\label{integral:eq5}
c\left(12\right)&=& \exp \left[-\beta \Phi\left(12\right)+\gamma\left(12\right)+
B\left(12\right)\right]-1-\gamma\left(12\right).
\end{eqnarray}
for the closure equation. Again, the RHNC approximation amounts to assume
$B(12)=B_0(r_{12})$, but clearly this is a much more drastic approximation
in the present case, as the real $B(12)$ depends on the relative
orientations of the particles whereas the reference $B_0(r_{12})$ does not.
As a result, one might expect the results to be less precise in this
case as compared with the isotropic fluid counterpart.

\subsection{Iterative procedure}
\label{subsec:iterative}
As discussed before, the iterative procedure in the case of the spherical
isotropic potential is very simple and outlined in Table I. It requires a
series of transformation to and from Fourier space, where the solution of
the OZ equation is more conveniently carried out in view of
Eq.(\ref{integral:eq3})
\begin{table}
{\large
\begin{equation*}
\begin{CD}
c\left(r\right)
@>\text{Fourier transform} >>
\hat{c}\left(k\right) \\
@AA\text{Closure } A  @.    @VV\text{OZ equation  }V\\
\gamma\left(r\right)
@<\text{Inverse Fourier transform} <<
\hat{\gamma}\left(k\right) \\
\end{CD}
\end{equation*}
}
\label{table:tab1}
\caption{Schematic flow-chart for the solution of the OZ equation
for isotropic potentials.}
\end{table}
The iterative solution of the angular dependent Ornstein-Zernike (OZ)
equation (\ref{integral:eq4}) along with the approximate closure equation
(\ref{integral:eq5}) again requires a series of direct and inverse Fourier
transforms between real and momentum space involving the bridge function
$B(12)$, the direct correlation function $c(12)$, the pair distribution
function $g(12)$, the total correlation functions $h(12)$ and the auxiliary
function $\gamma(12)=h(12)-c(12)$.

In addition to this, however, the orientational degrees of freedom
introduce additional direct and inverse Clebsch-Gordan (CG) transformation
between the coefficients of the angular expansions in different frames
\cite{Gray84}. Two important examples are the so-called ``axial'' or
``molecular'' frames, with $\hat{\mathbf{z}}= \hat{\mathbf{r}}_{ij}$ in
real space, and the $\{\mathbf{k}\}$ representation with $\hat{\mathbf{z}}
= \hat{\mathbf{k}}$ in momentum space, because in those representations
some of the calculations become particularly simple. This set of
transformations also allows the definition of the correlation functions (in
particular the $g(12)$) in an arbitrarily-oriented axes. We further note
that, in the presence of an anisotropic potential, the Fourier transform is
implemented through a Hankel transform, and that the cylindrical symmetry
of the angular dependence included in the Kern-Frenkel potential of
Section~\ref{sec:model} (when the number of patches present on each sphere
is one or two) allows us to use the simpler version of the procedure for
linear molecules.

All necessary equations can be found in Ref.~\onlinecite{Gray84}, that is
the standard reference for this topic, and only the most important ones
will be reported in the following.

The resulting scheme is illustrated in Table II and is the extension of the
isotropic case given in Table I~\cite{Giacometti09b}. 
\begin{table} 
{\large
\begin{equation*} 
\begin{CD}
c\left(r;l_{1}l_{2}l\right)
@>\text{Hankel transform }>>
\tilde{c}\left(k;l_{1}l_{2}l\right)
@>\text{Inverse CG transform }>>
\tilde{c}_{l_{1}l_{2}m}\left(k\right) \\
@AA\text{CG transform } A  @.    @VV\text{OZ equation  }V\\
c_{l_{1}l_{2}m}\left(r\right)
@.      @.
\tilde{\gamma}_{l_{1}l_{2}m}\left(k\right) \\     
@AA\text{Expansion inverse} A  @.    @VV\text{CG transform }V\\
c\left(r,\omega_1,\omega_2\right)
@.      @.
\tilde{\gamma}\left(k;l_{1}l_{2}l\right) \\
@AA\text{Closure } A  @.    @VV\text{Inverse Hankel transform }V\\
\gamma\left(r,\omega_1,\omega_2\right)   
@.      @.
\gamma\left(r;l_{1} l_{2} l\right) \\
@AA\text{Expansion } A   @.   @VV\text{Inverse CG transform  }V\\
\left[\gamma_{l_{1} l_{2} m}\left(r\right) \right]_{\text{old}}   
@<<<    \text{Iterate}    @<<<  
\left[\gamma_{l_{1} l_{2} m}\left(r\right) \right]_{\text{new}}
\end{CD}
\end{equation*}
}
\label{table:tab2}
\caption{Schematic flow-chart for the solution of the OZ equation
for the Kern-Frenkel angle-dependent potential.See Section \ref{sec:integral} for a description of the scheme}
\end{table}
Consider the expansion in spherical harmonics of the auxiliary function
$\gamma(12)$ in the axial frame
\begin{eqnarray}
\label{integral:eq6}
\gamma(12) &=& \gamma\left(r,\omega_1,\omega_2 \right) = 4\pi \sum_{l_1,l_2,m} \gamma_{l_1 l_2 m} 
\left(r\right) Y_{l_1m}\left(\omega_1\right) Y_{l_2\bar{m}}\left(\omega_2\right),
\end{eqnarray}
where $\bar{m} \equiv -m$, and its inverse reads
\begin{eqnarray}
\label{integral:eq7}
\gamma_{l_{1} l_{2} m} \left(r\right)&=&\frac{1}{4\pi} \int d\omega_1d\omega_2 \gamma\left(r,\omega_1,\omega_2 \right) 
Y_{l_{1}m}^*\left(\omega_1\right)
Y_{l_{2}\bar{m}}^*\left(\omega_2\right). 
\end{eqnarray}
Eq.(\ref{integral:eq7}) provides the initial set $[\gamma_{l_1 l_2
m}(r)]_{\text{old}}$ described as the initial point in Table II, whereas
Eq.(\ref{integral:eq6}) yields the next term in the iteration map
$\gamma(r_{12},\omega_1,\omega_2)$. 

Using the aforementioned RHNC closure approximation $B(12)=B_0(r_{12})$,
the bridge function is constructed and then inserted in
Eq.(\ref{integral:eq5}) to get $c(r,\omega_1,\omega_2)$.
Eq.(\ref{integral:eq7}) with the replacement $\gamma \to c$ is then
exploited to infer the corresponding axial coefficients $c_{l_{1} l_{2}
m}(r_{12})$

The next step is to implement a Clebsh-Gordan transform in direct space, in
order to transform from axial coefficients $c_{l_{1} l_{2} m}(r)$ where
$\hat{\mathbf{z}} = \hat{\mathbf{r}}$ to space coefficients $c\left(r;l_{1}
l_{2} l \right)$ in an arbitrarly oriented frame. The necessary expressions
are the equation pairs
\begin{eqnarray}
\label{integral:eq9}
c\left(r;l_{1} l_{2} l \right)&=& \left(\frac{4 \pi}{2l+1}\right)^{1/2}
\sum_{m} C \left(l_{1} l_{2} l ; m \bar{m} 0 \right) c_{l_{1} l_{2} m} \left(r\right),
\end{eqnarray}
where the $C \left(l_{1} l_{2} l ; m \bar{m} 0 \right)$ are Clebsch-Gordan
coefficients, with the inverse transform given by
\begin{eqnarray}
\label{integral:eq10}
c_{l_{1} l_{2} m} \left(r\right)&=& \sum_{l} C \left(l_{1} l_{2} l ; m \bar{m} 0 \right) \left(\frac{2l+1}{4\pi}\right)^{1/2} 
c\left(r;l_{1} l_{2} l\right).
\end{eqnarray}
and the coefficients $c(r;l_{1} l_{2} l )$ are then given by
Eq.(\ref{integral:eq10}).

The last required tool is the Fourier transform of the radial parts, that
is a Henkel transform as given by the pairs
\begin{eqnarray}
\label{integral:eq11}
\tilde{c}\left(k;l_{1} l_{2} l \right) &=& 4 \pi \mathrm{i}^{l}
\int_{0}^{\infty} dr~ r^2 c\left(r;l_{1} l_{2} l \right) j_{l} \left(k r\right),
\end{eqnarray}
with the inverse transform reading
\begin{eqnarray}
\label{integral:eq12}
c\left(r;l_{1} l_{2} l \right) &=&
\frac{1}{2 \pi^2 \mathrm{i}^l} \int_{0}^{\infty} dk~ k^2 
\tilde{c}\left(k;l_{1} l_{2} l \right) j_l \left(kr\right),
\end{eqnarray}
where $j_l(x)$ is the spherical Bessell function of order $l$. By using
``raising'' and ``lowering'' operations on the integrands these are finally
evaluated with $j_0(kr)=\sin kr/kr$ kernels, for $l$ even, and
$j_{-1}(kr)=\cos kr/kr$ kernels, for $l$ odd. Fast Fourier Transforms are
programmed for both instances. (Note that Hankel transforms
$\tilde{X}\left(k;l_{1} l_{2} l \right)$ are imaginary for $l$ odd while
for $l$ even they are real.)
 
Having obtained $\tilde{c}(k;l_{1} l_{2} l)$ from Eq.(\ref{integral:eq11}),
we have then reached the turning point in Table II, from which the
returning part can then be started with a parallel sequence of operation in
Fourier space. These include, a backward Clebsch-Gordan transformation to
return to an axial frame in $\mathbf{k}$ space and get
$\tilde{c}_{l_{1}l_{2} m}(k)$; a OZ equation in $\mathbf{k}$-space to get
$\tilde{\gamma}_{l_{1} l_{2} m}(k)$, followed by a forward Clebsch-Gordan
transformation, and an inverse Hankel transform, to find $\gamma(r;l_{1}
l_{2} l)$. A final backward Clebsch-Gordan transformation, brings a new
estimate of the original coefficients $[\gamma_{l_{1} l_{2}
m}(r)]_{\text{new}}$. This cycle is repeated until self-consistency
between input and output $\gamma_{l_{1} l_{2} m}(r)$ is achieved as before. 

Note that the OZ equation required in $\{\mathbf{k}\}$ representation,as
expressed in terms of the axial expansion coefficients of the transformed
pair functions, is given by the following matrix form
\begin{eqnarray}
\label{integral:eq13}
\tilde{\gamma}_{l_{1} l_{2} m}\left(k\right) &=& \left(-1\right)^{m} \rho
\sum_{l_{3}=m}^{\infty} 
\left[\tilde{\gamma}_{l_{1} l_{3} m}\left(k\right)+
\tilde{c}_{l_{1} l_{3} m}\left(k\right) \right]
\tilde{c}_{l_{3} l_{2} m}\left(k\right),
\end{eqnarray}

\subsection{Thermodynamics}
\label{subsec:thermodynamics}
Once that the correlation function $h(12)$ (and hence the distribution function $g(12)=h(12)+1$) is known, the
excess free energy can be computed as \cite{Lado82b,Giacometti09b}

\begin{eqnarray}
\label{integral:eq14}
\frac{\beta F_{\rm ex}}{N} &=& 
\frac{\beta F_{1}}{N} +\frac{\beta F_{2}}{N}+\frac{\beta F_{3}}{N},
\end{eqnarray}
where
\begin{eqnarray}
\label{integral:eq15a}
\frac{\beta F_1}{N} &=& -\frac{1}{2} \rho \int d \mathbf{r}_{12} 
\left \langle \frac{1}{2} h^2\left(12\right)+h\left(12\right)-g\left(12\right) \ln \left
[ g\left(12\right) e^{\beta \Phi\left(12\right)} \right] \right \rangle_{\omega_{1} \omega_{2}}, \\
\label{integral:eq15b}
\frac{\beta F_2}{N}&=& - \frac{1}{2 \rho} \int \frac{d\mathbf{k}}{\left(2\pi\right)^3}
\sum_{m} \left \{ \ln \mathrm{Det} \left[ \mathbf{I} + \left(-1\right)^m \rho
\tilde{\mathbf{h}}_m \left(k\right) \right] - \left(-1\right)^m
\rho \mathrm{Tr} \left[\tilde{\mathbf{h}}_m \left(k\right) \right] \right \}, \\
\label{integral:eq15c}
\frac{\beta F_3}{N}&=& \frac{\beta F_3^0}{N}-\frac{1}{2} \rho \int d \mathbf{r}_{12} \left \langle \left[ g\left(12\right)-g_0\left(12\right) 
\right] 
B_0\left(12\right) \right 
\rangle_{\omega_{1} \omega_{2}}.
\end{eqnarray}
In Eq. (\ref{integral:eq15b}), $\tilde{\mathbf{h}}_m(k)$ is a Hermitian matrix with elements $\tilde{h}_{l_1 l_2 m}(k)$, $l_1,l_2 \ge m$, and $\mathbf{I}$ is the unit matrix. The last equation, for $F_3$, directly expresses the RHNC approximation. Here $F_3^0$ is the reference system contribution, computed from the known free energy $F^0_{\rm ex}$ of the reference system as $F_3^0=F^0_{\rm ex}-F_1^0-F_2^0$, with $F_1^0$ and $F_2^0$ calculated as above but with reference system quantities. 
 
The bridge function $B_0(12)$ appearing in (\ref{integral:eq15c}) is the key approximation in the RHNC scheme, since it replaces the unknown bridge function $B(12)$ in the general closure equation (\ref{integral:eq5}). This is taken from the Verlet-Weiss expression of the hard-sphere model
\cite{Verlet72},
as anticipated, in view of its simplicity and of the fact that it works reasonably well for the case of the square-well as we will see, but with
a renormalized diameter $\sigma_0$ for the hard-sphere that is selected by enforcing the variational condition \cite{Lado82}
\begin{eqnarray}
\label{integral:eq16}
\rho \int d \mathbf{r} \left [ g_{000}\left(r\right) - g_{\rm HS}
\left(r;\sigma_0\right) \right]  \frac{\partial B_{\rm HS} \left(r;\sigma_0\right)}{\partial \sigma_0}&=&0.
\end{eqnarray}

From the free energy $F$, one of course compute all thermodynamics following standard procedures. For the computation of the
phase diagram, the pressure and chemical potential are required at any fixed temperature.

The virial pressure $P$ is obtained as \cite{Gray84}
\begin{eqnarray}
\label{integral:eq17}
P &=& \rho k_B T - \frac{1}{3V} \left \langle \sum_{i=1}^N \sum_{j>i}^N
r_{ij} \frac{\partial}{\partial r_{ij}} \Phi\left(ij \right)
\right \rangle =\rho k_B T - \frac{1}{6} \rho^2 \int d \mathbf{r}_{12}
\left \langle g\left(12\right) r_{12} \frac{\partial}{\partial r_{12}} \Phi\left( 12 \right)
\right \rangle_{\omega_{1} \omega_{2}}.
\end{eqnarray}
that in turns can be cast in the form involving the cavity function 
$y(12)=g(12) e^{\beta \Phi(12)}$ with the result \cite{Giacometti09b}
\begin{eqnarray}
\label{integral:eq18}
\frac{\beta P}{\rho} &=& 1+ \frac{2}{3} \pi \rho \sigma^3 \left \{ 
\left \langle y\left(\sigma,\omega_1,\omega_2 \right) 
e^{\beta \epsilon \Psi\left(\omega_1,\omega_2\right)}\right \rangle_{\omega_1 \omega_2}
-  \lambda^3 \left \langle y\left(\lambda \sigma,\omega_1,\omega_2 \right) \left[
e^{\beta \epsilon \Psi\left(\omega_1,\omega_2\right)}-1 \right] \right \rangle_{\omega_1
\omega_2} \right \},
\end{eqnarray}

As already remarked, one of the main advantages of the RHNC closure stems from the fact that the calculation of the chemical
potential does not introduce any approximation in addition to that already included in the closure. It can be obtained from the thermodynamic relation
\begin{eqnarray}
\label{integral:eq19}
\beta \mu = \frac{\beta F}{N} + \frac{\beta P}{\rho},
\end{eqnarray}
 that was already used in Eq.(\ref{thermodynamic:eq2}).

Finally, we note that the ideal quantities for the free energy, the virial pressure and the chemical potential are
\begin{eqnarray}
\label{integral:eq20}
\frac{\beta F_{\rm id}}{N} = \ln (\rho \Lambda^3)-1 &\qquad& \frac{\beta P_{\rm id}}{\rho}=1 \qquad \beta \mu_{\rm id} = \ln (\rho \Lambda^3)
\end{eqnarray}
\section{Barker-Henderson perturbation theory}
\label{sec:perturbation}
Another powerful method to access thermophysical properties of a fluid is thermodynamical perturbation theory, that directly extracts
the free energy of the system from the knowledge of the free energy $F_0$ of a reference fluid (the hard-sphere fluid in the present case).
This is a well-known techniques in several fields of physics, including simple \cite{Hansen86} and complex \cite{Gray84} fluids, and hinges
on the fact that often the system under investigation is not very different from the reference one, so that an expansion in this
perturbation term is a reasonable approximation.  
Under these conditions, the results are expected to be rather reliable, even when stopping the expansion at the lowest orders.

In the square-well fluid case the analysis has been carried out in details in the late 70s \cite{Barker76,Henderson71}, starting from
the pionieering work by Zwanzig \cite{Zwanzig54}, and recently extended to the patchy case \cite{Gogelein08,Gogelein11}. 

Working in the grand-canonical ensemble as this is the most convenient one \cite{Henderson71},  we assume the total potential $U$ to have the following form
\begin{eqnarray}
\label{perturbation:eq1}
U_{\gamma}\left(1,\ldots,N\right)&=& U_{0}\left(1,\ldots,N\right)+\gamma U_{I}\left(1,\ldots,N\right)\\ \nonumber 
&=&\sum_{i<j} \Phi_{\gamma}\left(ij\right) = \sum_{i<j} \Phi_{0}\left(ij\right)+\gamma \sum_{i<j} \Phi_{\text{I}}\left(ij\right)\,\mbox{,}
\end{eqnarray}
where $U_{0}(1,\ldots , N)=\sum_{i,j} \Phi_{0}(ij)$ is the unperturbed part and $U_{I}(1,\ldots ,N)=\sum_{i,j} \Phi_{\text{I}}(ij)$
is the perturbation part. Here $0\le \gamma \le 1$ is used as perturbative parameter. Note that when each coordinate $i$ includes both
the coordinate $\mathbf{r}_i$ and patch orientation $\hat{\mathbf{n}}_i$, so that  $i\equiv (\mathbf{r}_i,\hat{\mathbf{n}}_i)$, then
the expression is valid also for the Kern-Frenkel model \cite{Gogelein08,Gogelein11}. For simple fluids, then $i \equiv \mathbf{r}_i$ only.
Introducing the following short-hand notation 
\begin{eqnarray}
\label{perturbation:eq2}
\int_{1,\ldots,N} \left(\cdots \right) &\equiv & \int \left[\prod_{i=1}^N d \mathbf{r}_i \left \langle \left( \cdots \right) 
\right \rangle_{\omega_{i}} \right]
\end{eqnarray}
for the integration over all particle coordinates, the grand-canonical partition function 
\begin{eqnarray}
\label{perturbation:eq3}
{\cal Q}_{\gamma} &=& \sum_{N=0}^{+\infty} \frac{e^{\beta \mu N}}{N!\Lambda_T^{3N} } \int_{1,\ldots,N} e^{-\beta U_{\gamma}}=
e^{-\beta \Omega_{\gamma}}
\end{eqnarray}
(here $\Lambda_T$ is the de Broglie thermal wavelength, and $\Omega_{\gamma}$ is the grand-potential) 
can then be used to obtain an expansion of the Helmholtz free energy \cite{Henderson71}.
\begin{eqnarray}
\label{perturbation:eq4}
F_{\gamma} &=& F_0+ \gamma \left(\frac{\partial F_{\gamma}}{\partial \gamma} \right)_{\gamma=0}+
\frac{1}{2!} \gamma^2 \left(\frac{\partial^2 F_{\gamma}}{\partial \gamma^2} \right)_{\gamma=0}+ \cdots
\end{eqnarray}
that is valid for arbitrary $\gamma$.

Taking the derivative of $\ln {\cal Q}_{\gamma}$ at fixed chemical potential $\mu$, one has, using Eq.(\ref{perturbation:eq1})
\begin{eqnarray}
\label{perturbation:eq5}
\left[ \frac{\partial}{\partial \gamma} \ln {\cal Q}_{\gamma} \right]_{\mu} &=& \frac{1}{2} \int_{1,2} \frac{\partial}{\partial \gamma}
\left[-\beta \Phi_{\gamma}  \left(12\right) \right] \rho_{\gamma} \left(12\right)\,\mbox{,}
\end{eqnarray}
where
\begin{eqnarray}
\label{perturbation:eq6}
\rho_{\gamma} \left(1\ldots h\right)&=& \frac{1}{{\cal Q}_{\gamma}}  
\sum_{N=h}^{+\infty} \frac{e^{\beta \mu N}}{\left(N-h\right)!\Lambda_T^{3N} } \int_{1,\ldots,N} e^{-\beta U_{\gamma}}\,\mbox{.}
\end{eqnarray}

When $\gamma=1$, this yields the free energy correct to first order  in the expansion Eq.(\ref{perturbation:eq4}).

The second order correction is far more laborious. Indeed, the extension of this analysis involves higher orders correlation functions and this
forces additional approximations to come into play \cite{Henderson71,Barker76}, thus humpering its practical utility.   In 1967, Barker-Henderson
gave a much simpler recipe \cite{Barker67} that was found to be rather effective in predicting the phase diagram of the square-well fluid
\cite{Henderson71} and was recently extended to the Kern-Frenkel case \cite{Gogelein11}.

The final result for arbitrary angular dependent potential, correct to second-order, reads \cite{Gogelein11}
\begin{eqnarray}
\label{perturbation:eq7}
\frac{\beta F}{N} &=&  \frac{\beta F_0}{N}+\frac{\beta F_1}{N}+\frac{\beta F_2}{N}+ \ldots\,\mbox{,}
\end{eqnarray}

where
\begin{eqnarray}
\label{perturbation:eq8}
\beta F_{1} &=& \frac{1}{2} \rho N \int d \mathbf{r}g_{0} \left(r\right) 
\left \langle \beta \Phi_{\text{I}} \left(r,\Omega,\omega_{1},\omega_{2}  \right)
\right \rangle_{\omega_{1},\omega_{2}}
\end{eqnarray}
and
\begin{eqnarray}
\label{perturbation:eq9}
\beta F_{2}&=& -\frac{1}{4} k_B T \rho N \left(\frac{\partial \rho}{\partial P} \right)_0 \int d \mathbf{r}g_{0} \left(r\right) 
\left \langle \left[\beta \Phi_{\text{I}} \left(r,\Omega,\omega_{1},\omega_{2}  \right) \right]^2
\right \rangle_{\omega_{1},\omega_{2}}\,\mbox{,}
\end{eqnarray}

In the particular case if the Kern-Frenkel potential, one obtains \cite{Gogelein08,Gogelein11}

\begin{eqnarray}
\label{perturbation:eq10}
\frac{\beta F_1}{N} &=& \frac{12 \eta}{\sigma^3} \int d \mathbf{r}  g_0\left(r\right) \left \langle
\beta \Psi\left(12\right) \right \rangle_{\omega_1,\omega_2}\,\mbox{.}
\end{eqnarray}
and
\begin{eqnarray}
\label{perturbation:eq11}
\frac{\beta F_2}{N} &=& -\frac{6 \eta}{\sigma^3}  \left(\frac{\partial \eta}{\partial P_0^{*}} \right)_T \int d \mathbf{r} 
g_0\left(r\right) \phi_{\text{SW}}^2 \left(r\right)
\left \langle \left[\beta \Psi\left(12\right) \right]^2 \right \rangle_{\omega 1,\omega_2}\,\mbox{,}
\end{eqnarray}
Here $P_0^*=\beta P_0/\rho$ is the reduced pressure of the HS reference system and $g_0(r)$ the corresponding radial distribution function.

As before, the pressure and chemical potential can be derived from the reduced free energy per particle $\beta F/N$, 
using the exact thermodynamic identities
\begin{eqnarray}
\label{perturbation:eq12a}
\frac{\beta P}{\rho} &=& \eta \frac{\partial}{\partial \eta} \left(\frac{\beta F}{N} \right) \\
\label{perturbation:eq12b}
\beta \mu &=& \frac{\partial}{\partial \eta} \left(\eta \frac{\beta F}{N} \right)\,\mbox{.}
\end{eqnarray}

The phase diagram in the temperature-density plane can then be computed using a numerical procedure that will discussed in connections with 
Eqs.(\ref{calculation:eq1a}) and (\ref{calculation:eq1b}) below, as it is identical to the one used for integral equations.
\section{Calculation of the fluid-fluid coexistence curves for integral equation and pertubation theory}
\label{sec:calculation}
The common feature of integral equations and thermodynamic perturbation
theory is that one is able to obtain approximate expressions for both the
pressure $P$ and the chemical potential $\mu$ as a function of the
temperature $T$ and the density $\rho$. In the presence of a fluid-fluid
(gas-liquid) transition, both $P$ and $\mu$ will have well defined
dependence on $T$ and $\rho$ in the gas and liquid branches, but not in the
coexistence regions. Hence, in order to obtain the coexistence curves, and
hence the phase diagram in the temperature-density plane two procedures are
possible. First a graphical procedure where one plots the chemical
potential versus the pressure at a given fixed temperature, and seeks the
intersections between the gas and the liquid branches. An example of this
procedure in the case of the square-well potential was given in
Ref.~\onlinecite{Giacometti09a}. This procedure, however, is neither very
practical nor very precise, as it involves qualitative deduction of the
crossing points. Yet it can be used as a first preliminary estimate of a
more precise numerical calculations as follows.

For a fixed temperature $T$, one can then compute the pressure of the gas
(colloidal poor) phase $P_{g}$ and of the liquid (colloidal rich) $P_{l}$
phases, and the corresponding chemical potentials $\mu_{g}$ and $\mu_{l}$.
The fluid-fluid (gas-liquid) coexistence line then follows from a numerical
solution of a system of non-linear equations 
\begin{eqnarray}
\label{calculation:eq1a}
P_{g} \left(T,\rho_{g} \right)&=& P_{l} \left(T,\rho_{l} \right) \\
\label{calculation:eq1b}
\mu_{g} \left(T,\rho_{g} \right)&=& \mu_{l} \left(T,\rho_{l} \right)
\end{eqnarray} 
whose solutions are the gas $\rho_{g}$ and liquid $\rho_{l}$ densities
associated with the coexistence lines.  By plotting the resulting
$\rho_{g}$ and $\rho_{l}$ as a function of $T$ the coexistence curve can be
constructed in the region where the transition occurs.  This procedure will
be followed in the analysis of the Kern-Frenkel fluid-fluid phase diagram
for both integral equations (see Section \ref{subsec:rhnc}) and
thermodynamic perturbation theory (see Section \ref{subsec:evaluation}),
but can also be exploited for the computation of the fluid-solid
transition, as explained in Section \ref{subsec:fluid-solid}.

\section{Results}
\label{sec:results}
\subsection{Fluid-Fluid coexistence curves from RHNC integral equation}
\label{subsec:rhnc}
\begin{figure}[t]
\begin{center}
\includegraphics[width=10cm]{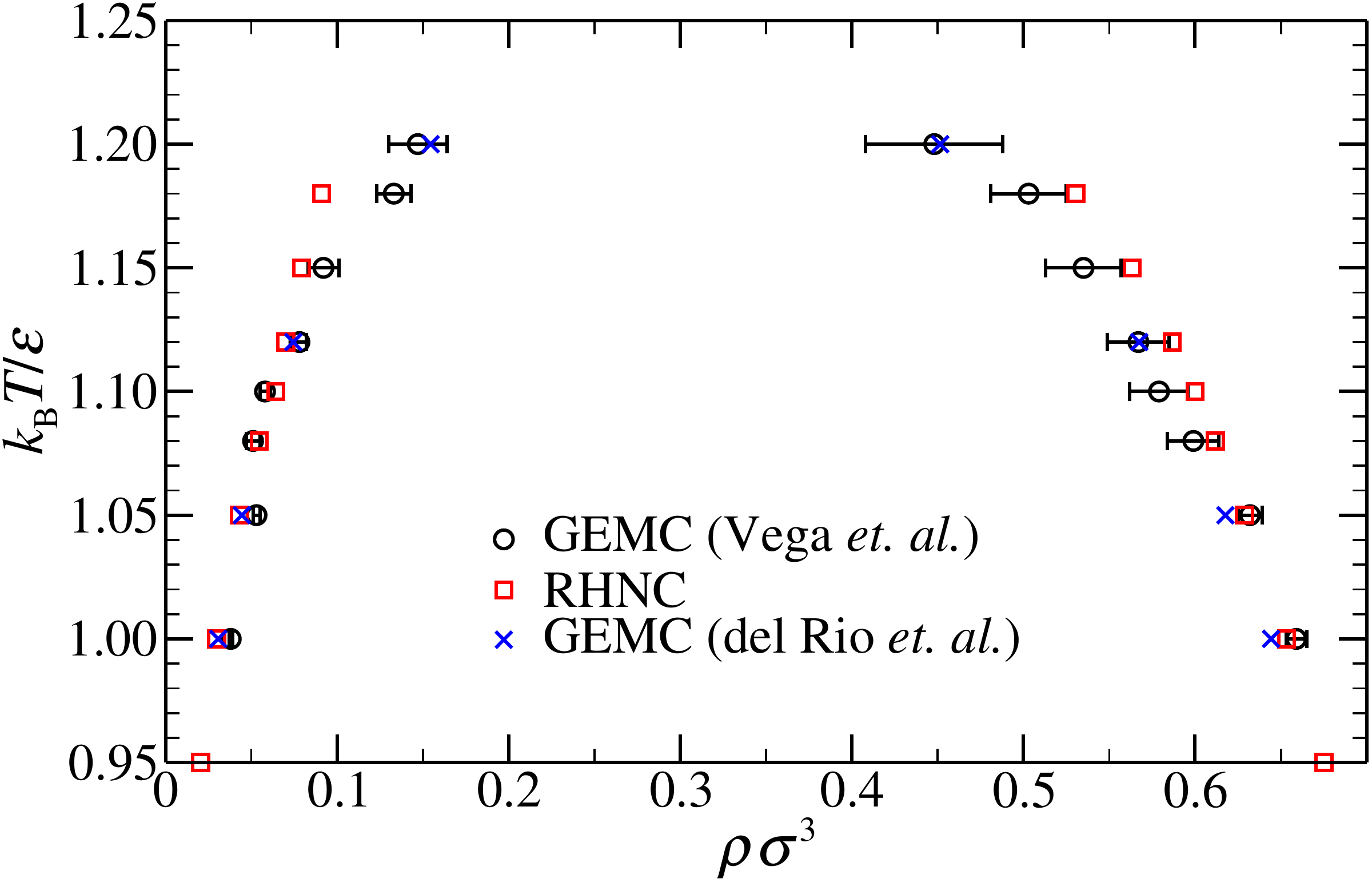}
\end{center}
\caption{Fluid-fluid coexistence curves $k_{\mathrm B} T/\epsilon$ versus
$\rho \sigma^3$ for the SW fluid of range $\lambda=1.5$. Data from RHNC
integral equation (squares) are contrasted with Monte Carlo simulations by
Vega et al.  \cite{Vega92} (circles), and del R\'{\i}o et al.
\cite{delRio02} (crosses). Adapted from Ref.\onlinecite{Giacometti09a}.
\label{fig:fig2}}
\end{figure}

We start by reviewing the quality of the RHNC integral equation theory for 
the simple case of the isotropic square-well case, a model which has often
be used as an effective potential (in the implicit solvent representation) to model colloidal particles interacting via depletion interactions\cite{noi1,noi2}.
Fig.\ref{fig:fig2} shows the RHNC predictions for the gas-liquid coexistence
from Ref.\onlinecite{Giacometti09a} contrasted with Monte Carlo simulations
on the same system carried out by different groups \cite{Vega92,delRio02}. 

Although the integral equation results appear to reproduce reasonably well
those from numerical simulations, two features are worth noticing.

First, thermodynamic inconsistencies associated with the approximate nature
of the closure are at the origin of the incorrect evaluation of the
pressure and chemical potentials and hence of the exact location of the
coexistence lines. This is an unavoidable feature of all integral
equations, and its origin and effects are well known. For most of the
cases, in fact, the whole critical region is inaccessible to integral
equation theories. 

A second additional point is related to the non-linear nature of the
self-consistence procedure, and gives rise to numerical instabilities that
may or may not be controlled depending on the considered state point. As a
general rule, lower temperatures and higher densities are harder to
converge, and hence for some points the solution of the system of Eqs.
(\ref{calculation:eq1a}) and (\ref{calculation:eq1b}) might not even exist.

\begin{figure}[t]
\begin{center}
\includegraphics[width=10cm]{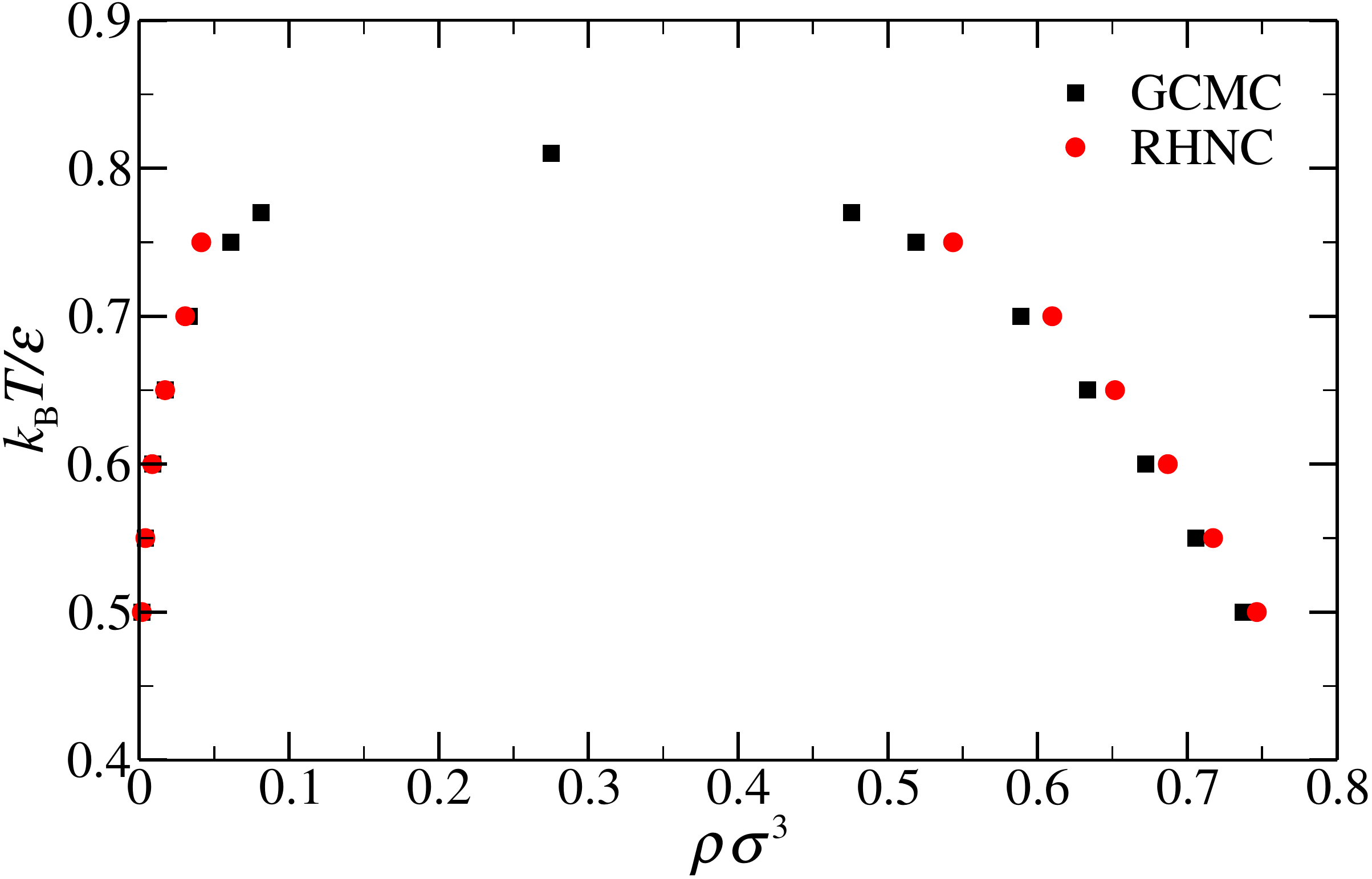}
\end{center}
\caption{Fluid-fluid coexistence curves $k_B T/\epsilon$ versus $\rho \sigma^3$ for the Kern-Frenkel fluid with coverage $\chi=0.8$ and range
$\lambda=1.5$. Data from RHNC integral equation (circles) are compared against MC simulations (squares). Adapted from Ref.\onlinecite{Giacometti09b}.
\label{fig:fig3}}
\end{figure}

These considerations are even more compelling in the more complex case of
the Kern-Frenkel fluid where condensation takes place at lower $T$
associated with the involved lower coverages. Hence one might expect an
agreement with respect to numerical simulations not better than what has
been reported for the isotropic square-well case. This is indeed the case,
as shown in Fig.\ref{fig:fig3} for the representative example of
single-patch particles with coverage $\chi=0.8$, for which $80\%$ of the
surface has a SW character, condensed into a unique patch. The width of the
square-well has been selected to be $\lambda=1.5$ as before, so that the
limit of full coverage gives back the result of Fig.\ref{fig:fig2}, and the
value $\chi=0.8$ has been selected to be half-way between the fully
occupied fluid and half-coverage that has peculiar behavior, as will be
discussed shortly.

In Fig.\ref{fig:fig3} we also report the coexistence lines and critical
points using Gibbs ensemble and Gran Canonical Monte Carlo simulations, as
outlined in Sections \ref{subsec:gibbs} and \ref{subsec:grand}. The former
were used to evaluate coexistence in the region where the gas-liquid
free-energy barrier is sufficiently high to avoid crossing between the two
phases, whereas the latter was used to locate exactly the critical point.
The details of the analysis can be found in Refs.
\onlinecite{Giacometti09a} (and in Ref.~\onlinecite{Giacometti10} for the
corresponding two-patch case).

The comparison between RHNC integral equation and MC simulations for the
Kern-Frenkel model with intermediate coverage $\chi=0.8$ displayed in
Fig.\ref{fig:fig3} shows that indeed the agreement is comparable with the
square-well case as anticipated. 

\begin{figure}[t]
\begin{center}
\includegraphics[width=10cm]{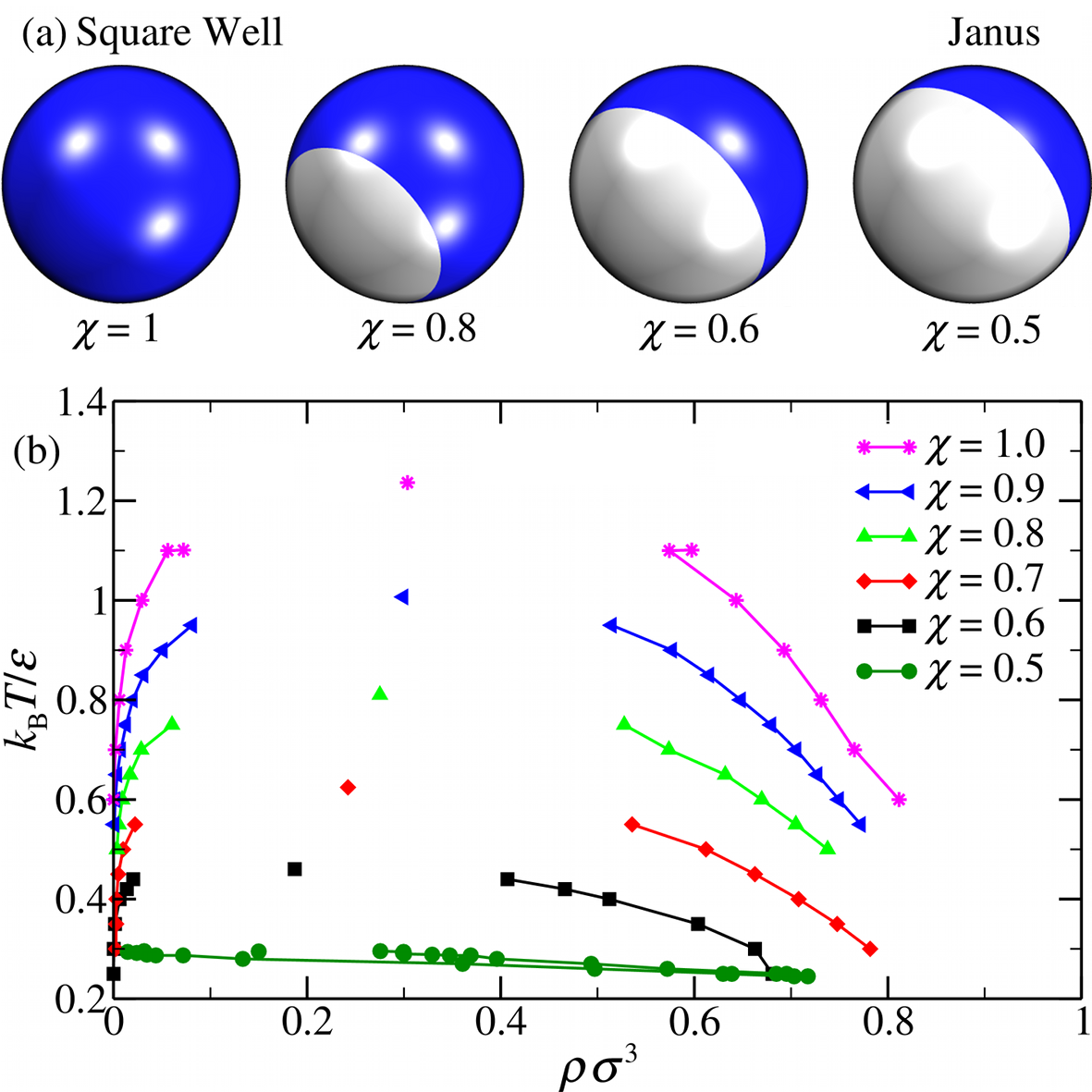}
\end{center}
\caption{(a) Cartoon of a one-patch particle with the coverage (fraction of
attractive surface, depicted in blue) changes from one (square well) to one
half (Janus). (b) Fluid-fluid coexistence curves $k_{\mathrm B} T/\epsilon$
versus $\rho \sigma^3$ for the Kern-Frenkel fluid with descreasing
coverages from $\chi=1.0$ (the SW case) to the Janus limit $\chi=0.5$. The
range is always set to $\lambda=1.5$. Data are from Gibbs ensemble MC
simulations for the coexistence lines and from grand-canonical MC
simulations for the critical points. This is the one-patch result: for the
two-patches counterpart, see Fig.\ref{fig:fig6}. Adapted from
Ref.\onlinecite{Sciortino10}.
\label{fig:fig4}}
\end{figure}

The Kern-Frenkel model offers the possibility to continuously change the
coverage interpolating from the isotropic square-well to the
symmetric Janus-like potential, when the coverage moves from $\chi=1$ to
$\chi=0.5$ (see Fig.~\ref{fig:fig4}-(a)). To investigate how the phase
diagram of Janus particles arises, we calculate how the gas-liquid
coexistence is modified on progressively reducing $\chi$.
Fig.~\ref{fig:fig4}-(b) shows MC simulations (Gibbs ensemble for the
coexistence lines and gran-canonical $\mu VT$ for the critical point)
results for the gas-liquid phase coexistence for several $\chi$ values,
extending the original data by Kern and Frenkel~\cite{Kern03}. A progressive
shrinking of the coexistence region to lower temperatures and densities
accompanies the decrease of the coverage. Consistently, as the coverage
decreases, both the critical temperature and critical density decrease (see
Fig.\ref{fig:fig7}). This is not surprising, in view of the fact that the
coverage is a measure of the attractive interactions intensity (by
controlling the maximum number of nearest neighbors
contacts~\cite{Bianchi06,23}) that dictates the value of $T_{\rm c}$ and
(perhaps more indirectly) $\rho_c$. It can be explicitly shown that this
shrinking of the coexistence region is a non-trivial one, and that it
cannot be inferred by a simple scaling of either the temperature or the
density. Up to $\chi=0.6$ coverage, however, the morphology of
the curve appears to be the standard one, with the gas and liquid
coexistence lines widening on cooling down.

The half-coverage $\chi=0.5$ is known as the Janus limit and will be
discussed in the next Section, as it displays an interesting unconventional
behavior \cite{Sciortino09}. 

\subsection{The Janus limit}
\label{subsec:janus}
The value of $\chi=0.5$ plays a special role in the Kern-Frenkel model with
a single patch. This can be seen by following the change in the coexistence
line locations as the coverage is reduced from the fully isotropic
square-well fluid ($\chi=1.0$) to the Janus case ($\chi=0.5$), as discussed
in the previous section.

\begin{figure}[t]
\begin{center}
\includegraphics[width=10cm]{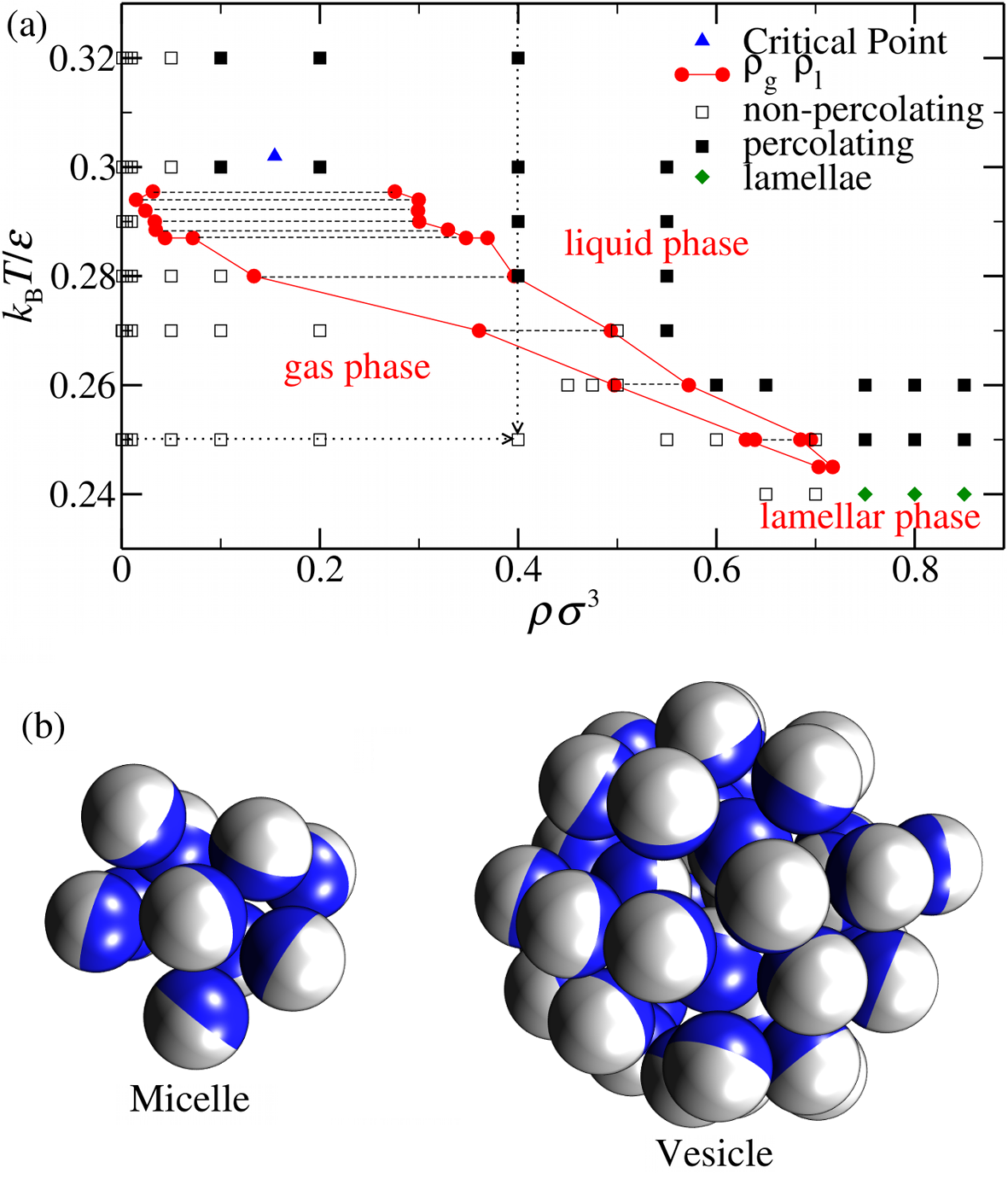}
\end{center}
\caption{(a) The anomalous phase diagram of the Janus limit ($\chi=0.5$),
adapted from Ref.\onlinecite{Sciortino09}. (b) Cartoon of the aggregates
which form in the gas phase on cooling, micelles and vesicles. The blue 
surface is attractive, modeled via a square-well potential.
\label{fig:fig5}}
\end{figure}

In Fig. \ref{fig:fig5} (a) the Janus case $\chi=0.5$, already reported in
Figure \ref{fig:fig4}, is shown by magnifying the very narrow region where
the transition occurs.

The difference with the standard phase diagram appears rather clearly. As
the fluid is cooled down to lower and lower temperatures, the coexistence
region shrinks, contrary to the standard case, and the two coexistence
lines appear to approach one another at sufficiently low temperatures. The
origin of this anomaly has far reaching consequences that have been
discussed in details~\cite{Sciortino09,Sciortino10}. The crucial point is
that at low temperature and density, monomers tend to aggregate in
micelles and vesicles (see Fig~\ref{fig:fig5}-(b)) bearing a well defined
number of particles ($10,41,\ldots$) so that all favorable contacts are
saturated inside the aggregate\cite{Sciortino09,Sciortino10}. This means
that, in all cases, the clusters always expose the hard-sphere parts to the
external fluid, and the condensation process is thus inhibited and
eventually destroyed altogether. The specificity of these magic numbers
can be also explained by means of a simple cluster theory~\cite{Fantoni11},
and an even simpler approach~\cite{Reinhardt11} can be developed to mimic
the onset of the re-entrant gas branch. Below the Janus limit, no evidence
of fluid-fluid transition was found, in full agreement with the above
interpretation~\cite{Sciortino09,Sciortino10}.

The unconventional shape of the phase diagram arises from the particular
energetic and entropic balance associated to the transition from the
micelles gas to the liquid phase. It has been found that, differently from
the usual gas-liquid behavior, the potential energy per particle is higher
in the liquid phase than in the micellar gas phase, a consequence of the
greater energetic stability of the micelles and vesicles as compared to the
disordered liquid phase. Hence, despite the fact that the gas phase is
stabilized by the translational entropy of the micelles, the coexisting
liquid phase is more disordered than the gas one. Such unconventional
entropic stabilization of the liquid phase arises from the orientational
entropy, since particles are orientationally disordered in the liquid phase
while they are properly oriented in the micelles gas phase.

\subsection{One versus two-patches}

\label{subsec:one}
It is interesting to investigate the effects of having the same attractive
square-well coverage split in two parts, at the opposite poles of the
sphere so that they are symmetrically distributed. This is the two-patche
case, and again this case smoothly interpolates between the fully occupied
isotropic SW case $\chi=1$ and the empty hard-sphere $\chi=0$ case.
\begin{figure}[t]
\begin{center}
\includegraphics[width=10cm]{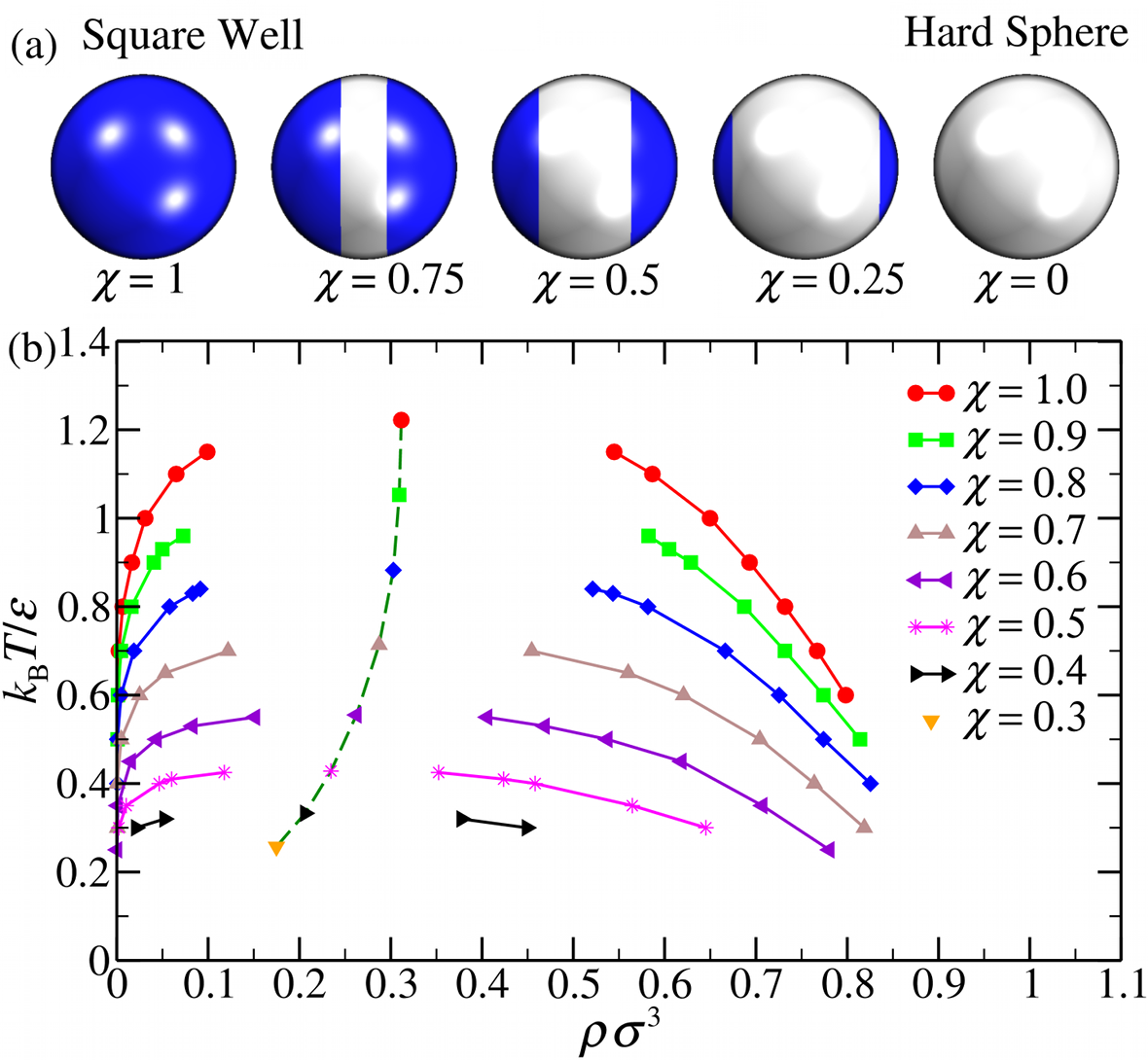}
\end{center}
\caption{(a) Cartoon of two-patch particles, when the coverage (blue)
varies from one (square well limit) to zero (hard sphere limit). (b)
Fluid-fluid coexistence curves $k_{\mathrm B}T/\epsilon$ versus $\rho \sigma^3$ for
the Kern-Frenkel fluid with two-patches and descreasing coverages from
$\chi=1.0$ (the SW case) to the value $\chi=0.3$. The range is always set
to $\lambda=1.5$. Data are from Gibbs ensemble MC simulations for the
coexistence lines and from grand-canonical MC simulations for the critical
points. This is the two-patches counterpart of Fig.\ref{fig:fig4}. Adapted
from Ref.\onlinecite{Giacometti10}.
\label{fig:fig6}}
\end{figure}
The corresponding fluid-fluid (gas-liquid) phase diagram is depicted in
Fig.\ref{fig:fig6}, that should be contrasted with the single patch
counterpart reported in Fig.\ref{fig:fig4}. Two main differences are
noteworthy. First, the case $\chi=0.5$ does not appear to play any
particular role, at variance with the single patch counterpart. Its
coverage value corresponds to the so-called tri-block Janus case, and will
be further discussed later on~\cite{Chen11,Romano11_a,Romano11_b}. This can
be attributed to the fact that at higher valences it is not possible to
saturate all favorable contacts, and hence the condensation process always
takes places in complete agreement with the previous interpretation of the
Janus case. The second main difference is related with the fact that in
the two-patch case coverages below $\chi=0.5$ do exhibit fluid-fluid
transition, as displayed in Fig.\ref{fig:fig6}, where coverages as low as
$\chi=0.3$ are depicted. Below this value, the transition becomes
metastable with respect to crystallization as explained in details in
Ref.~\onlinecite{Giacometti10}.

\begin{figure}[t]
\begin{center}
\includegraphics[width=10cm]{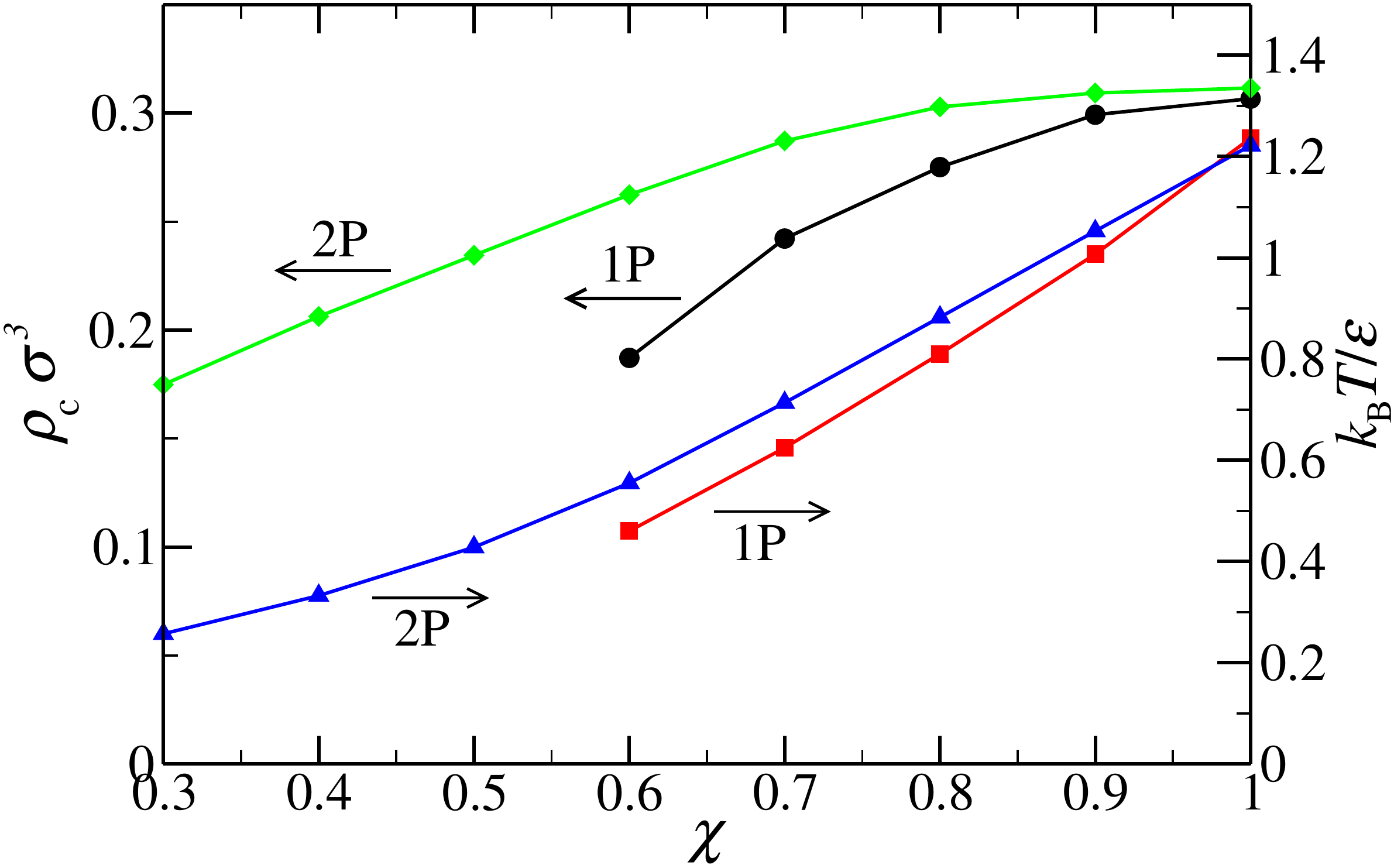}
\end{center}
\caption{Comparison between the one and the two-patches coverage dependence
of the reduced critical density (right axis) and of the reduced critical
temperature (left axis). Adapted from Ref.~\onlinecite{Giacometti10}.
\label{fig:fig7}}
\end{figure}
The $\chi$ dependence of the critical point can be inferred by plotting the
reduced critical densities and temperatures, as a function of the different
coverages, as shown in Fig.\ref{fig:fig7} \cite{Giacometti10}. Clearly, the
two-patch densities and temperatures lie always above the one-patch
counterparts, thus supporting the interpretation that it is easier to form
a fluid with higher valence at a given coverage, a feature that is related
to the existence of the fluid-fluid transition even for low coverages. 

\subsection{Evaluation of the fluid-fluid coexistence curves from
thermodynamic perturbation theory}
\label{subsec:evaluation}
As in the case of integral equation theory, the fluid-fluid phase diagram
for in the temperature-density plane can be computed even from
Barker-Henderson (BH) thermodynamic perturbation theory \cite{Gogelein11},
as indicated in Section \ref{sec:calculation}. 

\begin{figure}[t!]
\begin{center}
\includegraphics[width=10cm]{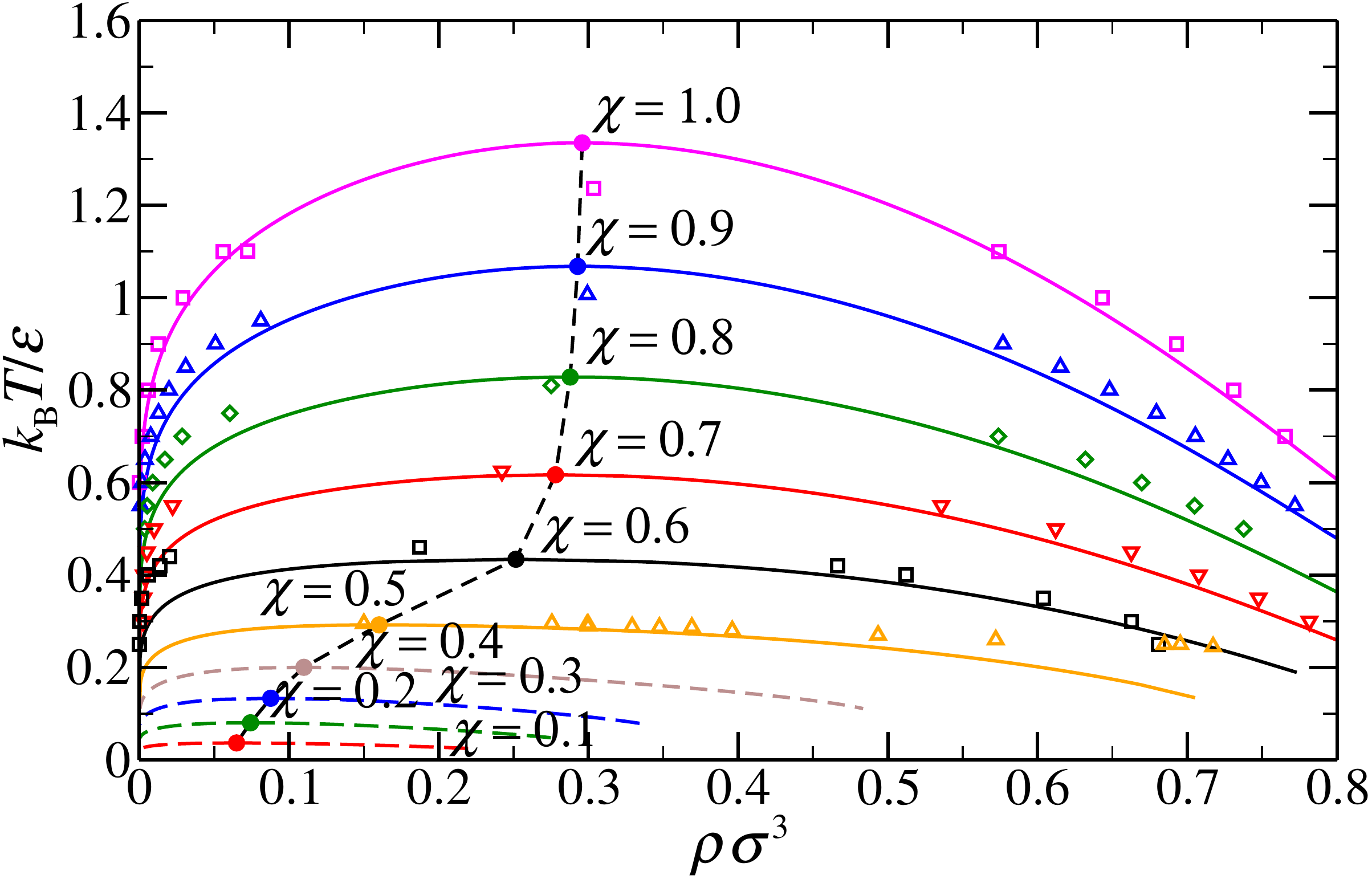}
\end{center}
\caption{The fluid-fluid phase diagram from BH thermodynamic perturbation
theory (continuous lines) contrasted against numerical simulations (open
symbols). Closed symbols refer to the BH critical points. Adapted from
Ref.\onlinecite{Gogelein11}.
\label{fig:fig8}}
\end{figure}

This is shown in Fig.~\ref{fig:fig8} for the one-patch case, and compared
with the same MC simulations used for comparison with integral equation
theory.

Given the simplicity of the theory, the accuracy of the BH theoretical
prediction is rather remarkable. On the one hand, this allows to provide a
prediction on the possible location of the true coexisting lines, even for
coverages where MC results are not yet available or in regions where
crystallization prevents the evaluation of the location of the metastable
gas-liquid critical point. Note that there are no restrictions in the
applicability of the BH theory, neither in thermodynamical parameters, nor
in coverages, the only limitations being the convergence of the numerical
non-linear solvers for Eqs.(\ref{calculation:eq1a}) and
(\ref{calculation:eq1b}). On the other hand, the BH perturbation theory
can be applied in its current form only to the one-patch case, as it lacks
the possibility of discriminating the location of the patches.
Notwithstanding these limitations, this approach is very promising as it
even allows the computation of the fluid-solid branch, as we will see in
the next section.

\subsection{The fluid-solid coexistence}
\label{subsec:fluid-solid}
\begin{figure}[t!]
\begin{center}
\includegraphics[width=10cm]{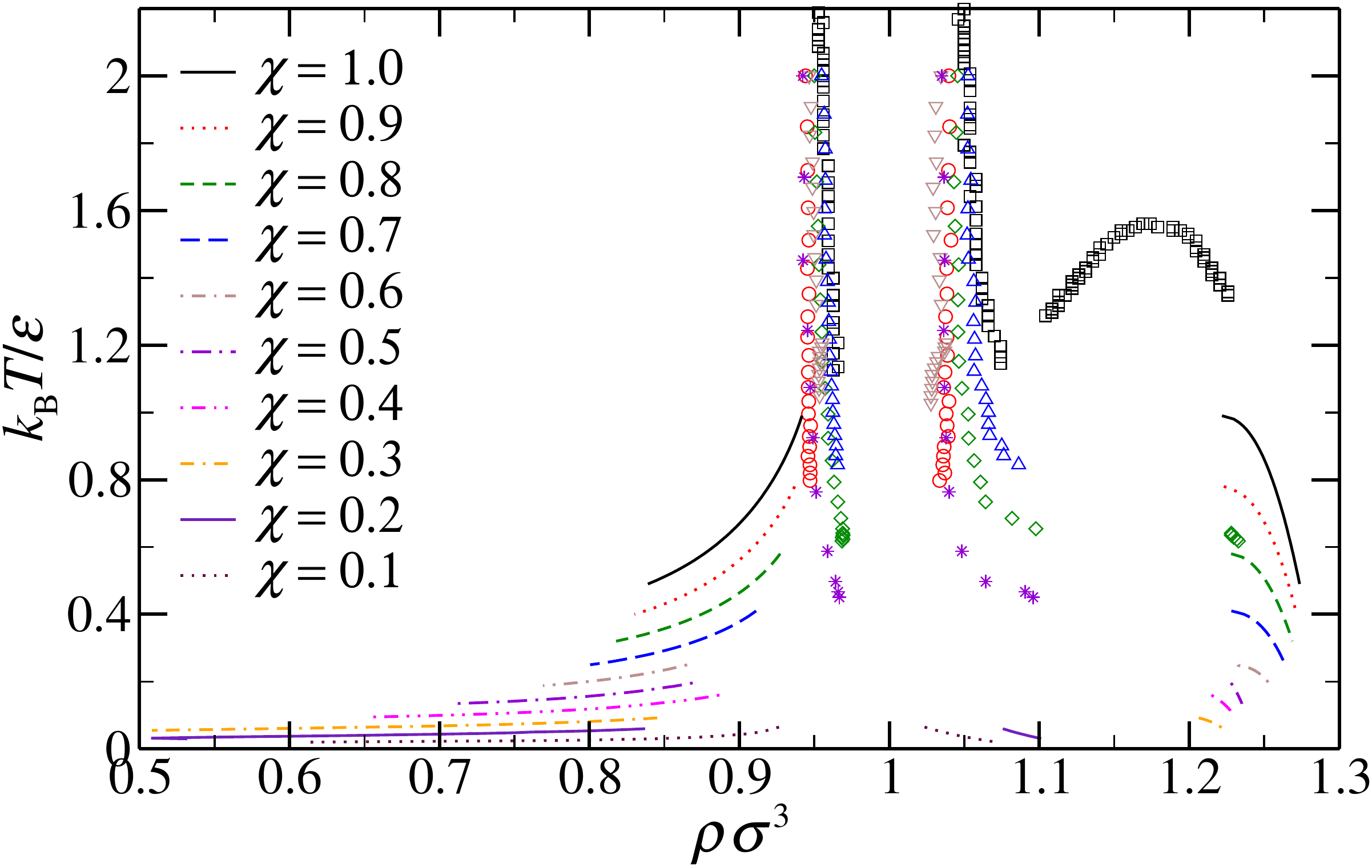}
\end{center}
\caption{The fluid-solid phase diagram from BH thermodynamic perturbation theory (continuos lines) contrasted
against numerical simulations (open symbols). Closed symbols refer to the BH critical points. 
The case $\chi=1.0$ (open squares) corresponds to the Young and Adler molecular dynamics results \cite{Young80} and includes the FCC-FCC transition
not considered in the patchy cases. Adapted from Ref.\onlinecite{Gogelein11}.
\label{fig:fig9}}
\end{figure}

At higher densities, a fluid-solid transition is expected in a way akin to
that occurring in the fully isotropic square-well case \cite{Young80,Liu05}.
For this value of the square-well amplitude $\lambda=1.5$, this transition
occurs at densities that are well-separated from the fluid-fluid
counterpart. In the isotropic SW case, this case has been studied several
times with different methodology \cite{Young80,Liu05}. In
Fig.\ref{fig:fig9} we report one of these study by Young and Adler
\cite{Young80} (open squares) carried out using molecular dynamics
techniques. In addition to the expected fluid-solid transition, with the
solid being a face-centered-cubic (FCC) lattice, an additional FCC-FCC
transition is visible as a large plateau at higher densities (see
Fig.\ref{fig:fig9}).
The other open symbols in Fig.\ref{fig:fig9} report results from the
fluid-solid transition of the Kern-Frenkel model with coverages from
$\chi=0.9$ down to the Janus limit $\chi=0.5$. In all cases, the final
crystal is a FCC with $12$ nearest-neighbors. There exists an additional
FCC-FCC transition akin to that found in the SW case, corresponding to a
transition from a more dilute to a denser FCC lattice, that is possible for
this range of the attractive well, but will not be considered here. Lines
in Fig.\ref{fig:fig9} report results from thermodynamic perturbation theory
obtained by using the same approach outlined above for the fluid-fluid
transition. In this case, a hint of the solid structural change is visible
at higher densities, but the exact coexistence lines could not be obtained
due to limitation in the convergence of the numerical algorithm associated
with the non-linear solver in Eqs.(\ref{calculation:eq1a}) and
(\ref{calculation:eq1b}). In spite of this drawback, thermodynamic theory
is able to capture the main features of the transition even for lower
coverages (as low as $\chi=0.1$). It should be stressed, however, that
transitions to different crystal structures may be envisaged at low
coverages, and this feature has not be accounted in the analysis reported
in Fig.~\ref{fig:fig9}.

\subsection{Self-assembly in a predefined Kagom\'e lattice}
\label{subsec:kagome}

The subtle interplay between the fluid-fluid and the fluid-solid transition
is one of the most delicate issue in the framework of self-assembly patchy
colloids, especially in the presence of additional effects such as
inhibiting clustering transitions. This was already hinted before, but a
very illuminating example of this is provided by the phase diagram of the
triblock Janus fluid, that has also been considered elsewhere in this
volume from the experimental point of view.

\begin{figure}[t!]
\centering
\includegraphics[width=0.95\textwidth]{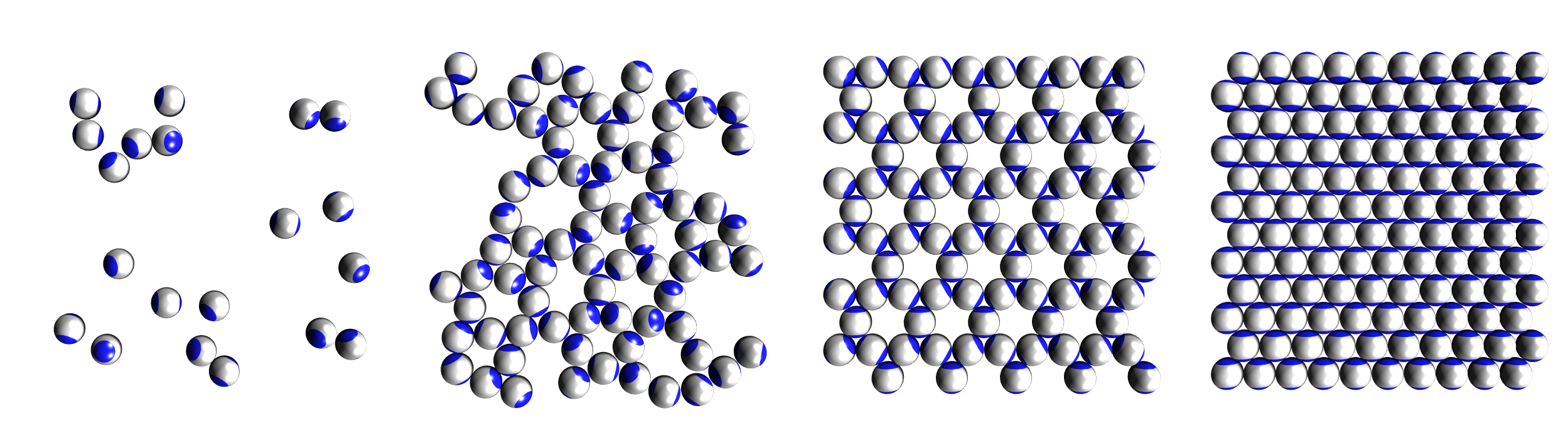}
\caption{\label{fig:fig10}
From left to right: snapshot of a gas, liquid, Kagome lattice and hexagonal
lattice. The Kagome and the hexagonal crystals are formed respectively at
low and high pressures.
Attractive patches are colored in blue, the hard--core remaining particle
surface is colored in gray. Particles are free to rotate in three
dimensions but are constrained to move on a flat surface. Adapted from Ref.\onlinecite{Romano11_b}.
}
\end{figure}

Triblock Janus colloids are spherical colloidal particles decorated with
two hydrophobic poles of tunable area, separated by an electrically charged
middle band (triblock Janus)~\cite{Chen11}. The electric charge of the
particles allows for a controlled switch of the interaction via addition of
salt, which effectively screens the overall repulsion, offering the
possibility to the hydrophobic attraction between patches to express
itself. Once deposited on a flat surface, after the addition of the salt,
particles organize themselves into a Kagome lattice. The crystallization
kinetics has been followed in real space in full details~\cite{Chen11}. The
patch width in the experimental system, of the order of 65 degrees
corresponding to $\chi \approx 0.57$, allows for simultaneous bonding of
two particles per patch, stabilizing the locally four--coordinated
structure of the Kagome lattice (see Fig.~\ref{fig:fig10}). 

\begin{figure}[t!]
\begin{center}
\includegraphics[width=10cm]{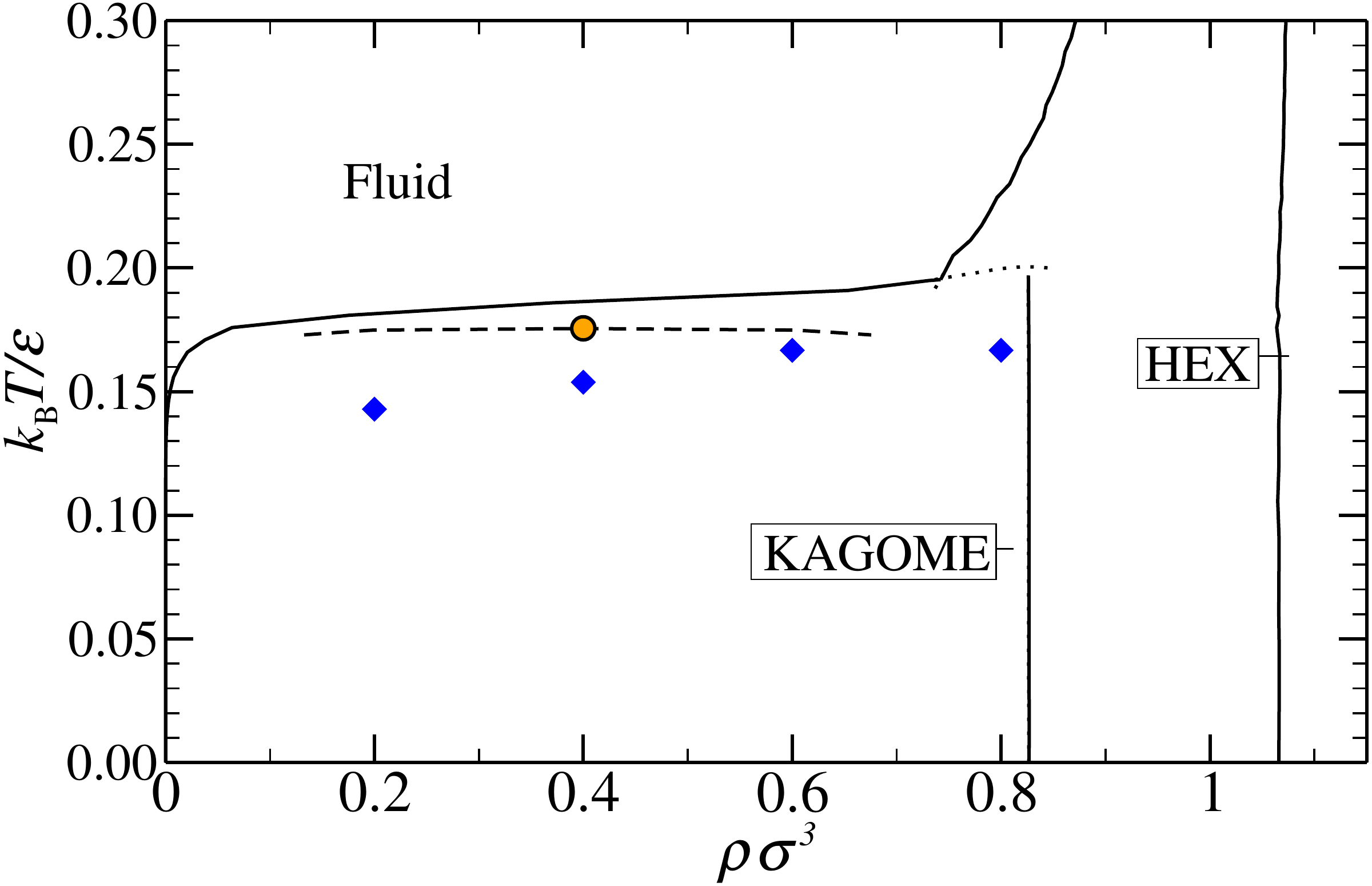}\\
\end{center}
\caption{
Phase diagrams in the $T-\rho$ for a wide ($\chi \approx 0.57$)
patch model (see Ref.~\onlinecite{Romano11_b} for the short range analog). Boundaries between stable phases are drawn as solid black
lines and metastable phase boundaries are dotted. The orange points in
 indicate the (metastable) gas--liquid critical point.
The dashed line represents the metastable gas--liquid phase
separation. Blue diamonds and red circles indicate the highest temperature
at which spontaneous crystallization into the Kagome and hexagonal lattice,
respectively, was detected at the corresponding pressure or density. Crosses
indicate the coexistence points checked via direct coexistence simulations. Adapted from Ref.\onlinecite{Romano11_b}.
\label{fig:fig11}}
\end{figure}

A triblock Janus particle can be modeled via the two-patches Kern-Frenkel
model 
in which the attractive region is split in two parts at the opposite poles
of the sphere, whereas the repulsive part is concentrated in the middle
strip at the equator. The square-well mimics the short-range hydrophobic
attraction, while the hard-sphere region represents the repulsive
charge-charge interaction. Depending on the considered state point and on
the patch amplitude several possible phases are possible, as shown in
Ref.~\onlinecite{Romano11_b} and summarized in Fig.\ref{fig:fig11}. In
full agreement with the experiments, we observe at comparable interaction
strength the spontaneous nucleation of a Kagome lattice. We also find at
larger pressure spontaneous crystal formation in a dense hexagonal
structure. Such easiness to crystallize suggests that in this system
crystallization barriers are comparable to the thermal energy at all
densities. Interestingly enough, we find that for this model a (metastable)
gas--liquid phase separation can be observed for large patch width.

The ability of accurately describing Janus triblock particles with the KF
potential is particularly rewarding and provides a strong support for the
use of such model for predicting the self--assembly properties of this
class of patchy colloids. The possibility of numerically exploring the
sensitivity of the phase diagram to the parameters (patch width and
interaction range) entering in the interaction potential provides an
important instrument and a guide to the design of these new particles to
obtain specific structures by self--assembly.


\section{Conclusions and open perspectives}
\label{sec:conclusions}
In this Chapter we have discussed some of the main theoretical approaches
that have been proposed in the last few years to address the computation of
the phase diagram in the temperature-density plane of the Kern-Frenkel
model, one of the paradigmatic models for patchy colloids. Three different
approaches, namely Monte Carlo simulations, integral equations and
perturbation theory have been discussed, and their performances contrasted
against each others. We have attempted to present the main ideas behind
each techniques, along with some representative examples of applications.
The main emphasis has been placed on the evaluation of the fluid-fluid
(gas-liquid) and fluid-solid phase diagrams, and their importance in the
framework of the self-assembly processes. As a matter of fact, the exact
location of the transition lines is a crucial ingredient necessary to
implement a bottom-up engineering of materials of new generation. It is
remarkable that none of these techniques are really new, being part of the
background given in any standard graduate course in Statistical Physics.
Yet, their implementation in the framework of patchy colloids often (if not
always) has required significant improvements that have shed new lights on
the techniques themselves. This is true for the newly improved Monte Carlo
schemes that have been devised in this field, but also for integral
equation and perturbation theories that have a long and glorious history.

Several steps need still to be performed. We still need an accurate study
of the role of the interaction range in selecting the most stable
geometries. More specifically, we need to understand under which patch
width and range conditions micelles and vesicles are the stable structures.
Experiments\cite{Hong08} indicate that for Janus particles one dimensional
structures becomes preferred when the interaction range is only a few
percent of the particle size. We also need to develop an accurate
methodology to predict all possible crystal structures for different types
of patterned particles (including Janus), their stability fields and the
associated nucleation rates. Understanding self-assembly into ordered
pre-defined structures can indeed have relevant industrial applications.
The Janus paradigm (and its theoretical counterpart, the Kern-Frenkel
model), can play an important role in this process.

The fact that notwithstanding its simplicity the Kern-Frenkel model is able
to present such a complex and rich scenario in its phase diagram is related
to the combined effect of two features. On the one hand, the short-range
and reversible nature of the involved interactions. This allows a partial
rearrangement, within a localized region in space, of the particles in
search of the optimal minimal energy configuration, a feature that would
not be possible for stronger (irreversible) covalent longer-range
interactions. On the other hand, the specificity dictated by the patchy
anisotropy is expedient in avoiding multiple degenerate configurations with
similar energies, thus eliminating defects and polidispersity effects
characteristic of isotropic colloids.

Qualitative and semi-quantitative agreements with experiments can be
obtained with the Kern-Frenkel model in some particular cases, and it is
hoped that the contribution of this book, including both theory and
experiments in a well chosen balance, will contribute to further strengthen
this very promising route.

\begin{section}*{Acknowledgments}
The results presented in this Chapter have been obtained in collaboration
with a number of people, including Christoph G\"ogelein, Fred Lado, Julio
Largo and Giorgio Pastore.
\end{section}

\bibliographystyle{apsrev}



\end{document}